\documentclass[10pt,english]{article}
\pdfoutput=1 
\usepackage{amsmath}
\usepackage{natbib}
\usepackage{amssymb}
\usepackage{babel}
\usepackage{graphicx}
\usepackage[margin=1in]{geometry}
\usepackage{caption}
\usepackage{subcaption}
\usepackage{epstopdf}

\author{
 Matthew Ricci\\
Brown University \\
  \texttt{matthew$\_$ricci$\_$1@brown.edu}
  \and
 Junkyung Kim\\
Brown University\\
  \texttt{junkyung$\_$kim@brown.edu}
 \and
Fredrik Johansson\\
Lund University\\
  \texttt{fredrik.johansson@med.lu.se}
}

\title{\noindent \rule{16cm}{3pt} \newline \begin{center} \textbf{A Passage-of-time  Model of the Cerebellar Purkinje Cell} \end{center} \noindent \rule{16cm}{.4pt}}
\date{}

\usepackage{hyperref}

\begin{document}
\maketitle

\begin{abstract}
\noindent The cerebellar Purkinje cell controlling eyeblinks can learn, remember and reproduce the interstimulus interval in a classical conditioning paradigm. Given temporally separated inputs, the cerebellar Purkinje cell learns to pause its tonic inhibition of a motor pathway with high temporal precision so that an overt blink occurs at the right time. Most models relegate the Purkinje cell's passage-of-time representation to afferent granule cells, a subpopulation of which is supposedly selected for synaptic depression in order to make the Purkinje cell pause. However, granule cell models have recently faced two crucial challenges: 1) bypassing granule cells and directly stimulating the Purkinje cell's pre-synaptic fibers during training still produces a well-timed pause, and 2) the Purkinje cell can reproduce the learned pause, invariant to the temporal structure of probe stimulation. Here, we present a passage-of-time model which is internal to the Purkinje cell and is invariant to probe structure. The model accurately simulates the Purkinje cell learning mechanism and makes testable electrophysiological predictions. Importantly, the model is a numerical proof-of-principle for a biological learning mechanism which does not rely on changes of synaptic strength. 
\end{abstract}

\section{Introduction}
While long-term potentiation (LTP) and long-term depression (LTD) are often measurable during memory encoding, the prevailing doctrine that such Hebbian changes in synaptic strength entirely explain memory has been challenged by several recent findings. In both \textit{Aplysia} \citep{Chen2014} and mammalian hippocampal cells \citep{Ryan2015}, long-term memory does not in all cases require the persistence of changes in synaptic strength. The result that deviates the furthest from Hebbian theory comes from a recent result in eyeblink conditioning, which showed that a single Purkinje cell can memorize a temporal relationship of hundreds of milliseconds between two inputs \citep{Johansson2014}.

\begin{figure}[h!]
\centering
\includegraphics[width=.5\textwidth]{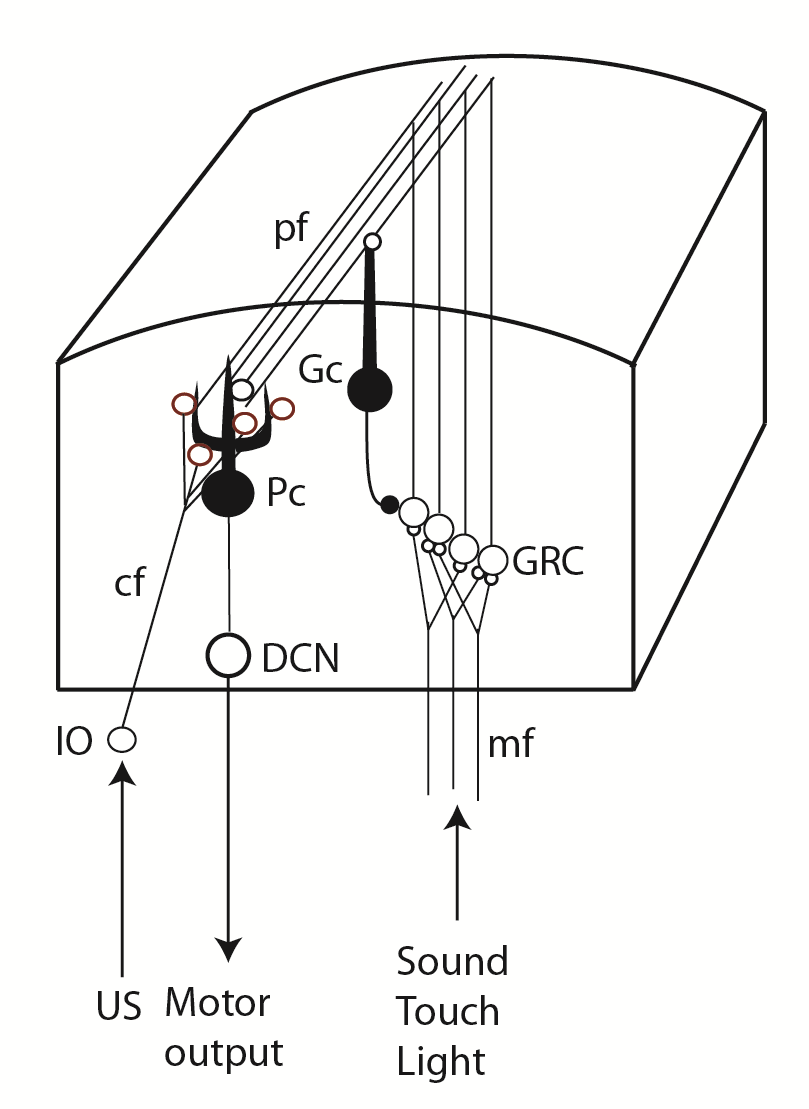}
\caption{\emph{Simplified neural circuitry}. The CS is transmitted via the mossy/parallel fiber system and the US via the climbing fiber. Key: IO: inferior olive. cf: climbing fiber. mf: mossy fibers. Pc: Purkinje cell. Gc: Golgi cell. pf: parallel fibers. GRC: granule cells. DCN: deep cerebellar nuclei.}
\label{circuit}
\end{figure}

If a conditional stimulus (CS), such as a tone, is repeatedly followed by a blink-eliciting unconditional stimulus (US), with a fixed temporal delay (the interstimulus interval or ISI), a blink response to the CS develops. This conditioned response (CR) occurs just before the expected US onset for ISIs from $\sim$100 ms to seconds \citep{Gormezano1969, Kehoe2002}. The underlying learning occurs in a specific microzone within the C3 zone of cerebellar cortical lobule HVI \citep{Yeo1985,Hesslow1994,Hesslow1994a,Mostofi2010,Heiney2014}. Because many and diverse conditional stimuli may predict the same unconditional stimulus, the association-forming mechanism should exhibit extensive fan-in, which the Purkinje cell indeed does; its vast dendritic arbor gets CS input from approximately 200,000 parallel fibers \citep{Harvey1991} that come from the tiny and densely packed granule cells (GRC) in the granular layer of the cerebellar cortex, while it gets US input from a single climbing fiber \citep{Mauk1986,Steinmetz1989,Hesslow1999,Hesslow2002} (Fig. \ref{circuit}) that contacts the entire arbor \citep{Eccles1967}.
\newline
\indent Purkinje cells, which are the sole output of the cerebellum, have high tonic firing rates due to an internal pacemaker mechanism \citep{Cerminara2004}. Given CS-US pairings, blink-controlling Purkinje cells learn an adaptively timed pause response \citep{Jirenhed2007, Halverson2015, Hesslow1995,Hesslow1994b,Jirenhed2011a} in their tonic inhibition of a motor pathway \citep{Hesslow2002}. This learned pause mirrors the known features of behavioral CRs. It is extinguished during repeated CS alone presentations and it is rapidly re-acquired \citep{Jirenhed2007}. The pause is always adaptively timed, reaches its maximal amplitude just before the predicted onset of the US, and ends shortly after, even if the CS lasts only a few milliseconds or outlasts the ISI by hundreds of milliseconds \citep{Jirenhed2011, Johansson2014}.  In temporal uncertainty paradigms where mixed trials of different ISIs are used, multiple temporally locked pause responses to the same CS are learned \citep{Halverson2015}. Importantly, the behavioral lower limit of a minimal ISI of $\sim$100 ms holds at the level of the single Purkinje cell \citep{Wetmore2014}.
\newline
\indent Contemporary modeling is dominated by the idea of the granular layer network generating a passage-of-time (POT) representation. \citep{Buonomano1994,Medina2000,Yamazaki2009}. Before learning, the Purkinje cell is assumed to fire at its tonic rate due to net excitation from the balanced activity of excitatory granule cells and inhibitory basket/stellate interneurons \citep{Medina2000a}.  In response to a CS drive from the mossy fibers, time-variance in granule cell (GRC) activity is assumed to arise from fast and partly lateral feedback inhibition from Golgi cell interneurons. The effect would be a series of random transitions between activity and quiescence in each granule cell, creating instantaneous population vectors that represent time. Spike-timing dependent plasticity then selectively recruits LTD for those GRC-to-Purkinje cell synapses most activated by the CS around the time of US onset. The parallel fiber-Purkinje cell synapse is anti-Hebbian in that correlated excitatory inputs from parallel fibers and the climbing fiber causes LTD. LTD tips the balance of afferent activity towards net inhibition near expected US onset, so that, after learning, the CS alone initiates a well-timed pause in Purkinje cell firing. 
\newline 
\indent Granule cell POT models cannot readily be reconciled with two recent results, one physiological and one computational. On the physiological side, Johansson and colleagues showed that the Purkinje cell itself learns the ISI \citep{Johansson2014}. They bypassed the granular layer network by delivering the CS directly to the parallel fibers. Purkinje cells learned well-timed pause responses to ISIs of 150, 200 and 300 ms.  Furthermore, both GABA-ergic interneurons \citet{Johansson2014} and glutamatergic AMPA receptors \citet{Johansson2015a} could be blocked without disrupting the pause response. Their result implies that the precisely timed Purkinje cell activity depends upon an internal timing mechanism that measures and stores temporal duration. 
\newline
\indent The computational process embodied by this internal mechanism deviates significantly from that of previous POT models. After training, the Purkinje cell can produce a well-timed pause given only the initial ($< 20 $ ms) part of the CS \citep{Svensson1999,Jirenhed2011}. For example,\citet{Johansson2014} found that drastically varying the CS on probe trials (17.5-800 ms, 100-400 Hz) had no effect on the learned pause. Existing POT models, on the other hand, require the probe stimulus to endure for at least as long as the training stimulus. For, as soon as the probe terminates, afferent activity would shift to its pre-stimulation conditions and return the net input of the Purkinje cell to excitation, preventing pause expression. The real computational process underlying the CR must contain a strong non-linearity which switches the Purkinje cell from a tonic spiking state to a pause-expressing state. 
\newline
\indent Hence, we see two crucial criteria for the modeling of a Purkinje cell timing mechanism:
\newline

\noindent \textbf{Physiological: } The neural machinery for pause learning and expression must be contained in the Purkinje cell itself.
\newline

\noindent \textbf{Computational: } The computational process initiating pause expression must be gated by a strong non-linearity or ``switch''. 
\newline

\noindent This second criterion has new neurobiological underpinning in the work of \citet{Johansson2015a}, who found that nanoliter injections of antagonists of the inhibitory metabotropic glutamate receptor 7 (mGluR7), targeted at the Purkinje cell dendrites, eliminates the pause response. This suggests that mGluR7 elicits the response through activation of protein-activated, hyperpolarizing K$^{+}$ channels in the family Kir3 or through inactivation of cAMP or Ca$^{2+}$ signaling. In other words, pause expression may be gated by a metabotropic switch. 
\newline
\indent Though this hypothesis may explain what triggers pause expression, it does not explain the temporal structure of the pause itself or how this structure is learned. An actual trained Purkinje cell receiving a parallel fiber signal, produces a pause in firing with a slightly delayed onset, accurately timed maximum and critically timed offset. \citet{Johansson2014a} provided a conceptual theory for how such a timing mechanism could work. They suggested a selection mechanism whereby CS onset prompts the Purkinje cell to release a batch of evolving ``recorder units''. These units could be second messengers, some unknown molecular switches, polynucleotides or evolving proteins. The precise biophysical identity of recorder units does not substantially affect the theory of \citet{Johansson2014a}; the only crucial feature of these units is that they evolve through different time-encoding states beginning at CS onset. Each unit contributes to inhibition of the cell at the interval that it encodes, and the strength of inhibition at any given time since CS is encoded by the quantity of units that encode the same interval, which increases via repeated exposure to CS-US pairs. Recorder units thus constitute an intracellular POT representation. Importantly, this model satisfies the two above criteria: the pause mechanism is internal to the Purkinje cell and it is mediated by a glutamatergic switch. 
\newline
\indent Below, this theory is instantiated as a computational model with two parts: a ``write'' module, which intracellularly records the interstimulus interval and a ``read'' module, which translates the archived intervals into a timed hyperpolarization event. These states become active only when afferent activity switches the cell from its passive OFF state to its active ON state. We trained this model cell with several paradigms typical to the eyeblink conditioning literature and found:
\begin{itemize}
\item Given ISIs ranging from 100 ms to 500 ms, the cell learns a pause response with critically timed onset, maximum and offset, as in \citet{Johansson2014}. 
\item After training, the cell produces the learned pause given only the initial part of the CS, as in \citet{Svensson1999}, \citet{Jirenhed2011} and \citet{Johansson2014}.
\item The learned pause is extinguished by CS-only trials in about as much time as initial acquisition, as in \citet{Jirenhed2007,Johansson2014}. 
\item Alternating ISIs across trials teaches the cell two pauses, as in \citet{Jirenhed2007,Johansson2014}. 
\end{itemize}

\noindent Additionally, we ran two simulations, which, to our knowledge, have not been tested on the biological Purkinje cell. First, we tried stimulating the cell with two-part CSs or two-part USs on each trial: stimuli consisted of two impulse trains to the same fiber separated by a period of silence, so that the cell was effectively exposed to two ISIs. In these simulations, our model cell only learns one ISI. Next, we tried varying the intertrial interval (ITI) along with the ISI. We found that time to pause acquisition in all simulations was a function of both of these temporal quantities. The behavior of this formal model may help to inform future electrophysiology.

\section{Model}
\begin{figure}[h!]
\includegraphics[width=\textwidth]{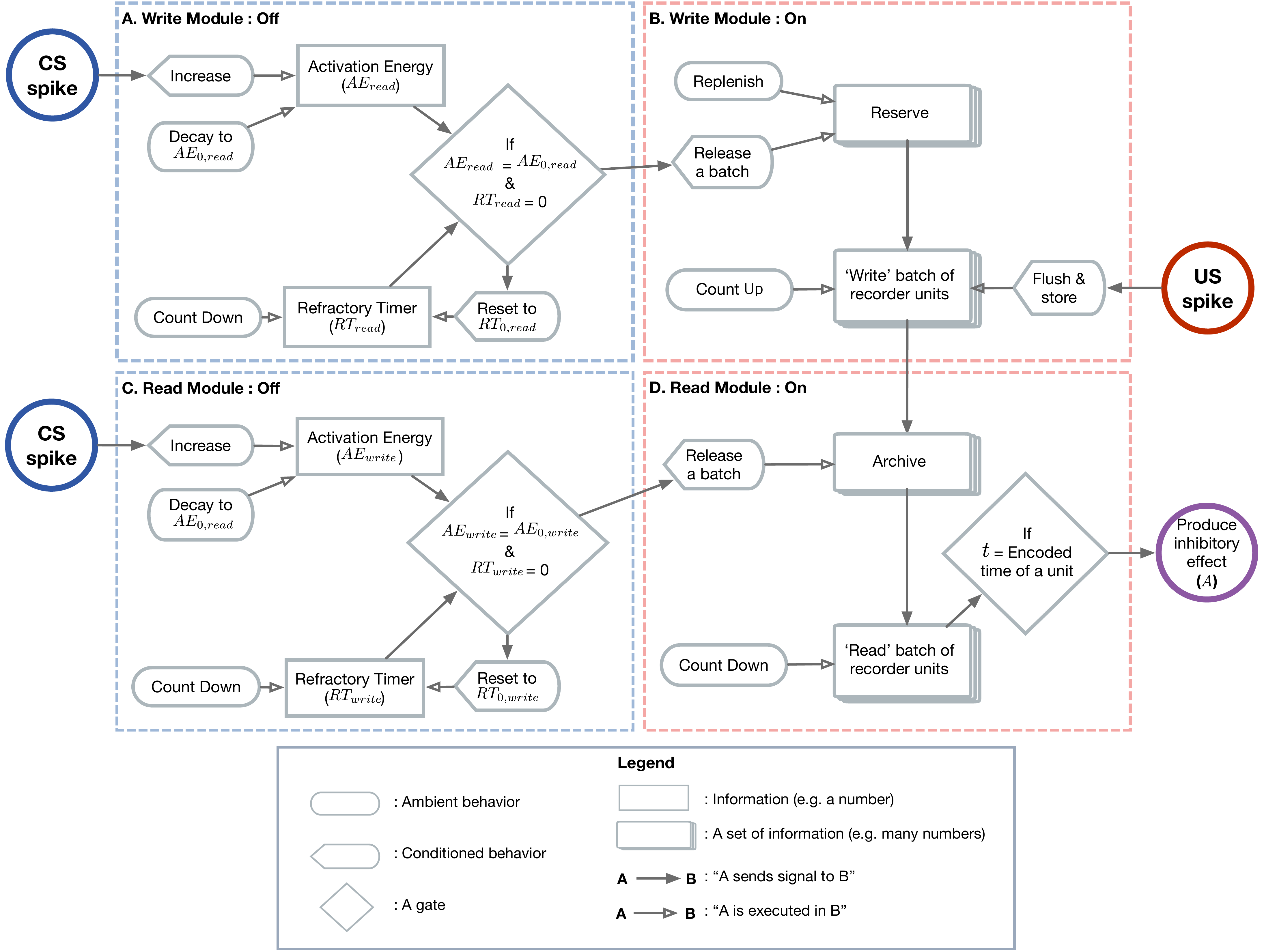}
\caption{\footnotesize\emph{Model diagram}. This diagram depicts the flow of events leading to the CR in the model Purkinje cell. \emph{Write/OFF)} The write module's switch is controlled by the quantity $AE_{\text{write}}$ (upper box). $AE_{\text{write}}$ is increased with each CS spike (blue circle) and is otherwise subject to a passive decay. Only when $AE_{\text{write}}$ reaches a threshold (diamond) does the cell transition to the write/ON state. When the switch is activated, a refractory timer (lower box) is reset and begins to tick downward until it reaches 0. \emph{Write/ON)} When the write module switches ON, the reserve (tiled boxes, top) continually ejects a batch of recorder units (tiled boxes, bottom). The reserve is then replenished until at maximum capacity. The ejected recorder units advance in state with the passage of time until the first US spike (red circle), at which point the batch is stored in the archive. \emph{Read/OFF)} The switch on the read module behaves exactly as in the write case, but with different parameter values. \emph{Read/ON)} A constant fraction of the archive (tiled boxes, top) is sampled. Each unit ``counts down" until the time encoded by its state. At this time (diamond), the unit introduces a fixed inhibitory current to the membrane (purple circle).}
\label{model}
\end{figure}

\emph{Model structure}. The Purkinje cell has two computational goals. First, it must record or ``write'' an ISI into memory. Second, it must ``read'' this ISI into a well-timed pause. As stated above, the Purkinje cell only transitions from its passive spiking behavior to its active read/write behavior when some type of switch is activated. Thus, we may formally model the cell as having two modules (read/write; Fig. \ref{model}, top vs. bottom row) which can be in two states (OFF/ON; Fig. \ref{model}, left vs. right column).
\newline
\indent Before CS stimulation, both the read and write modules are OFF and the cell spikes at its tonic rate. When pre-synaptic spikes arrive at the cell, a quantity called ``activation energy'' ($AE_{\text{write}}$ and $AE_\text{read}$) begins to increase. When $AE_{\text{write}}$ reaches a threshold value, the write module switches to the ON state. A refractory timer on this switch is then reset so that the cell is insensitive to the CS for a given period. 
\newline
\indent When the write module is activated after sufficient CS stimulation, a batch of ``recorder units'' is released from a ``reserve''. These units are a formal abstraction meant to stand for whatever process encodes the POT within the Purkinje cell. The reserve is a way of imagining these units before the write module is activated. After some recorder units have been released from the archive, the reserve begins to replenish at a constant rate. The released recorder units evolve through states in a way that mirrors the passage of time. For example, if they can encode four states, $A = 50$ ms, $B = 100$ ms, $C = 200$ ms, and $D = 300$ ms, then the POT is represented by the sequential evolution of the recorder units $A\to B \to C \to D$. Units in the reserve encode 0 ms. 
\newline
\indent At US onset, this evolution stops and the batch of recorder units is stored in an ``archive''. Again, this archive is simply a way of imagining the units after their evolution has been terminated by US onset (unless US onset is too early, for reasons discussed below). We assume that the evolution of the batch is noisy, so that, when the evolution stops, there are some units in each state. Thus, the stored batch is a histogram which constitutes an estimate on the ISI. For example, if the ISI is 200 ms, then there will be some units in states $A$, $B$, and $C$. We choose to imagine the recording medium as consisting of many individual ``units'', since we want to use a stored histogram (which displays the frequencies of many elements or samples) to estimate the ISI. The routine of write/OFF $\to$ write/ON $\to$ store batch is repeated with each CS presentation, as long as they are spaced far enough apart to respect the refractory timer on the write module. With each presentation, a batch of recorder units is released, evolves and is stored in the archive, updating the total number of recorder units in the archive encoding each time value. 
\newline
\indent Parallel to the writing process is a reading process, which is itself gated by a switch mechanism (Fig. \ref{model}, bottom left). Whether the read switch is activated depends on the read module's own activation energy ($AE_\text{read}$) and refractory timer. When $AE_\text{read}$ reaches a threshold, the read module transitions to the ON state (Fig. \ref{model}, bottom right). The activated reading module samples a batch of stored recorder units from the archive. Each recorder unit in this batch contributes a fixed amount of inhibition to the cell at a time corresponding to its state. For instance, if batch is released from the archive at time $t=0$, then a unit in state $A$ will introduce inhibition to the membrane potential at $t = 50$ ms. Note that, the more units there are in a batch encoding a particular time, the greater the inhibition will be at this time. In this way, the shape of the archive's histogram is directly translated into the timecourse of inhibition.
\newline
\begin{figure}[h!]
\begin{subfigure}{0.5\textwidth}
\includegraphics[width=0.9\linewidth, height=5cm]{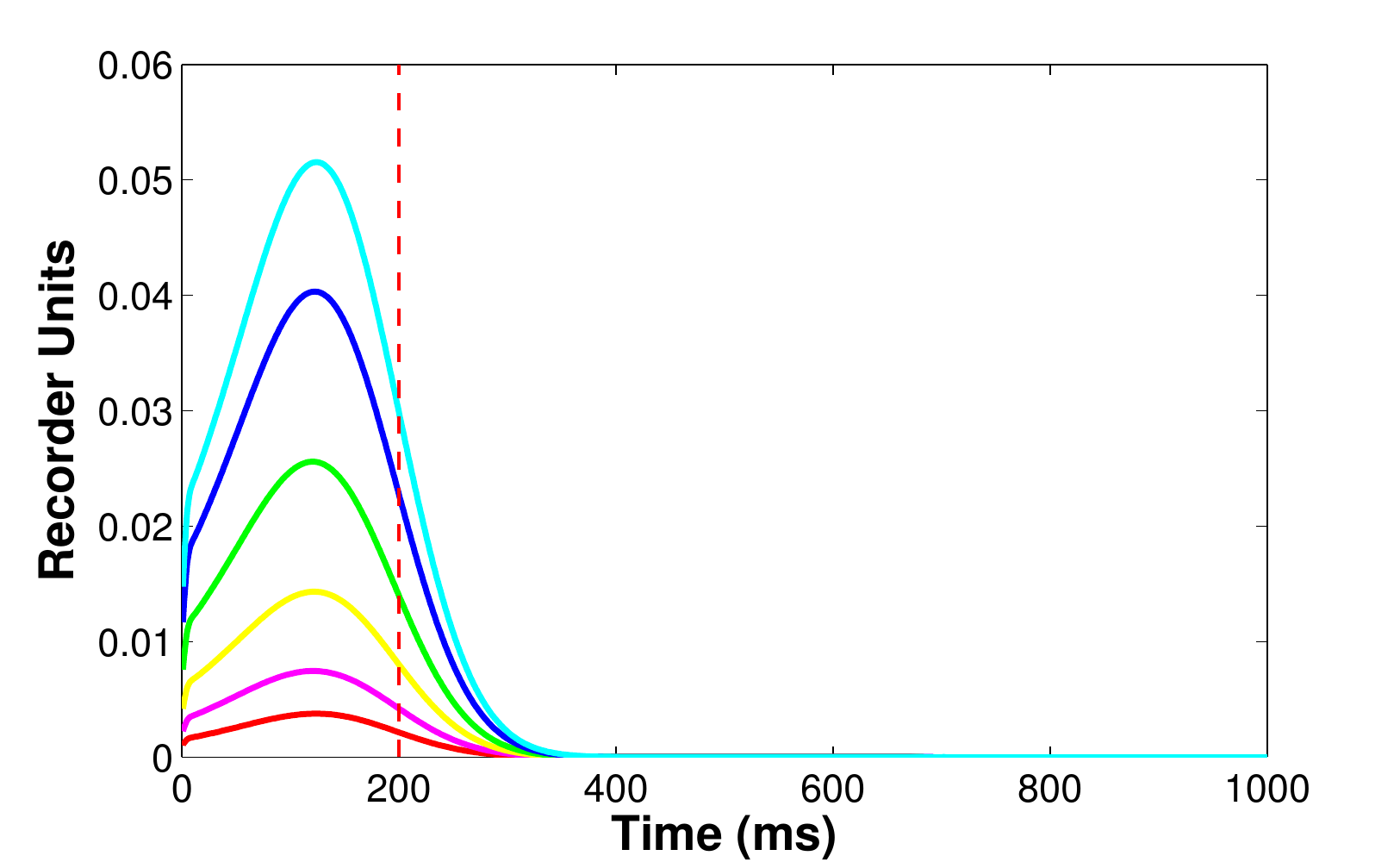} 
\caption{ISI = 200 ms}
\label{archives1}
\end{subfigure}
\begin{subfigure}{0.5\textwidth}
\includegraphics[width=0.9\linewidth, height=5cm]{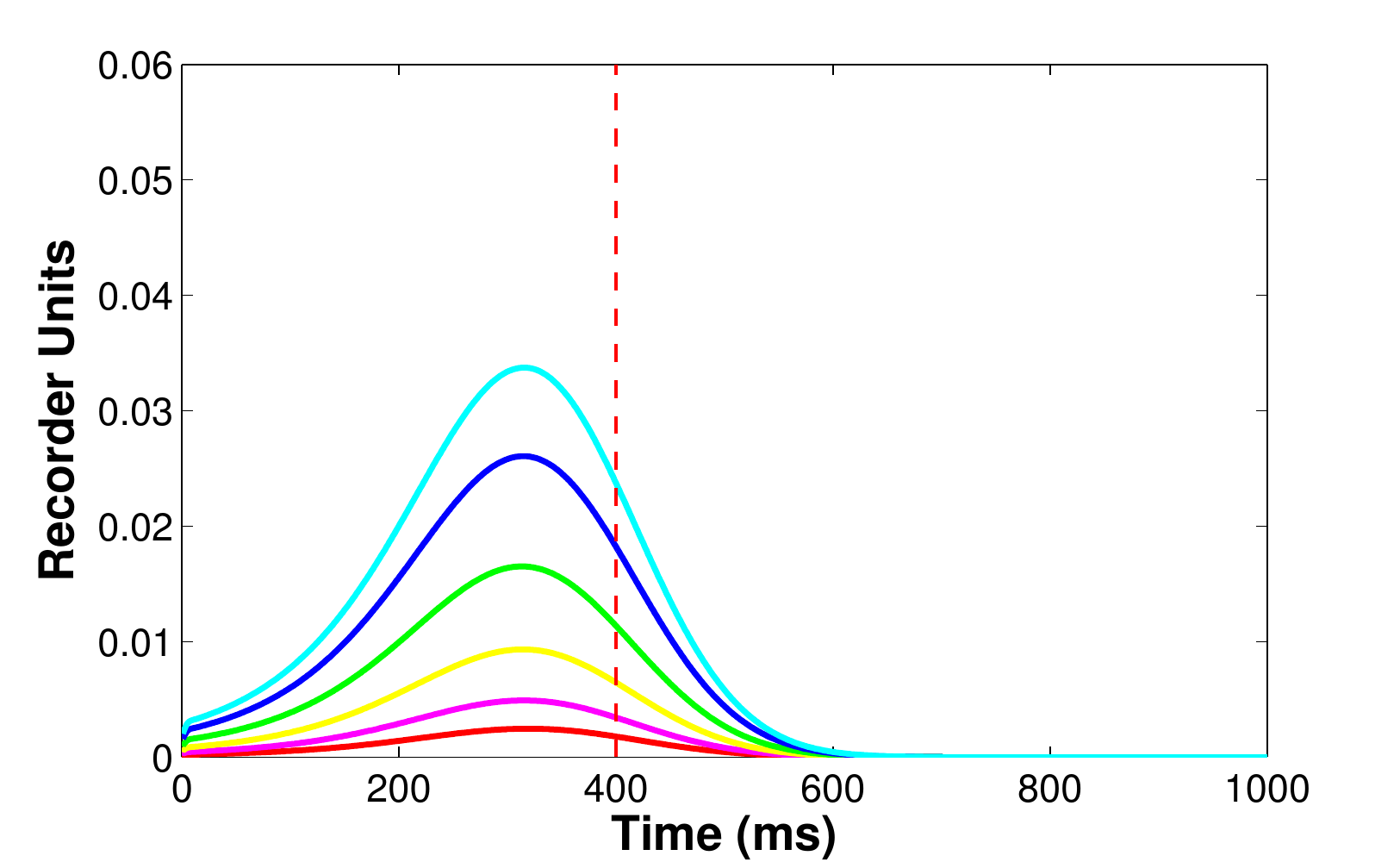}
\caption{ISI = 400 ms}
\label{archives2}
\end{subfigure}
\begin{subfigure}{.5\textwidth}
\includegraphics[width=0.9\linewidth, height=5cm]{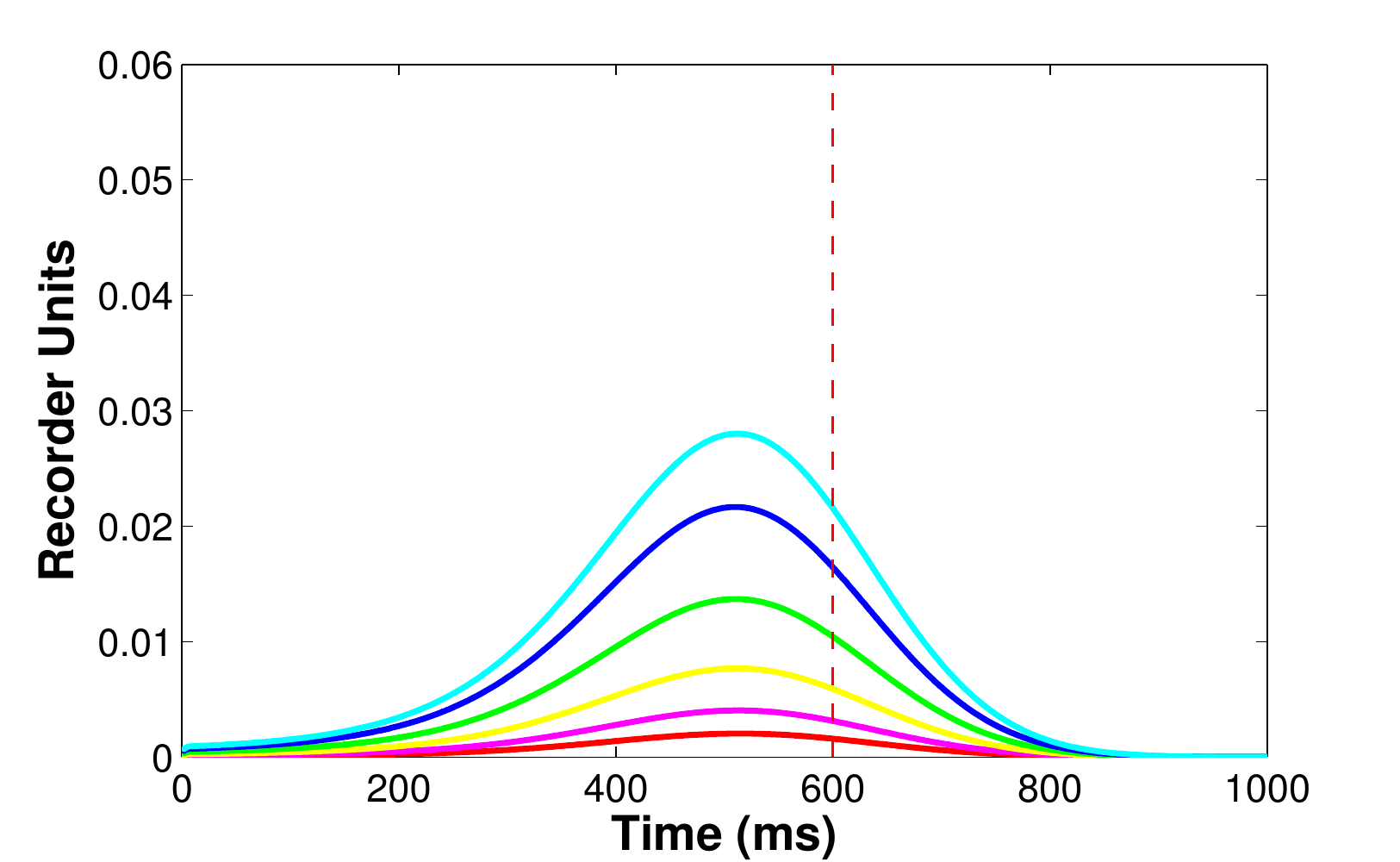}
\caption{ISI = 600 ms}
\label{archives3}
\end{subfigure}
\begin{subfigure}{0.5\textwidth}
\includegraphics[width=0.9\linewidth, height=5cm]{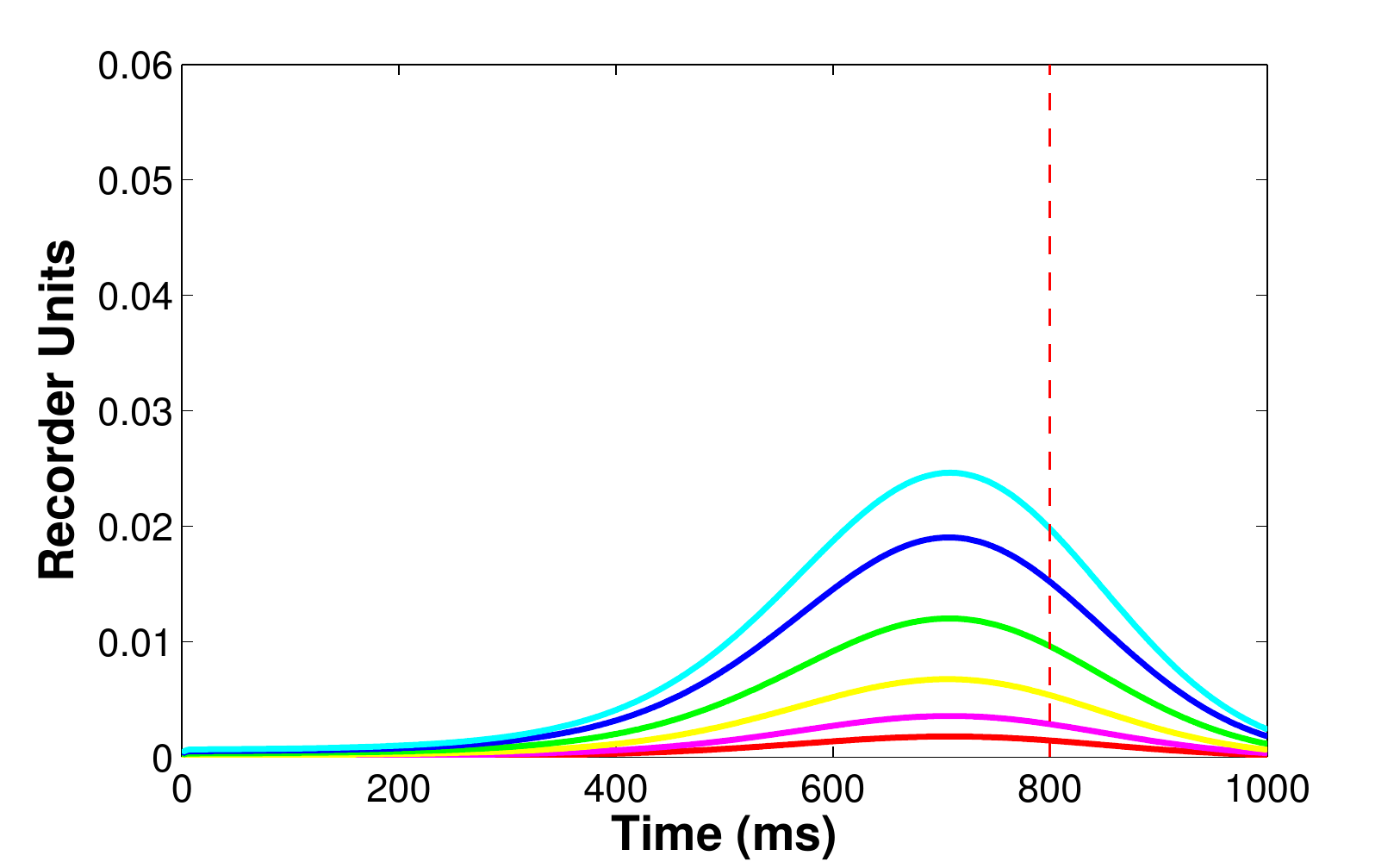}
\caption{ISI = 800 ms}
\label{archives4}
\end{subfigure}
\caption{\footnotesize \emph{ISI and number of trials determine archive shape.} Each curve plots the archive ($y-$axis, number of units encoding each time step on the $x-$axis) for different training durations and ISIs. Colors indicate 20 (red), 40 (magenta), 80 (yellow), 160 (green), 320 (blue), 640 (cyan) training trials. Further training trials make the archive grow taller. Panels indicate 200 (top left), 400 (top right), 600 (bottom left) and 800 (bottom right) ms ISIs. Longer ISIs make the archive less peaked. These different shapes translate into different inhibition profiles via the reading module.}
\label{archives}
\end{figure}

\noindent \emph{Model details}: We model the Purkinje cell as a single-compartment, leaky integrate-and-fire-neuron (Eq. \ref{mempot}) whose parameters are given in Table \ref{cellparams}. This simple neural spiking model is appropriate for our purposes, since the Purkinje cell CR is internally generated and therefore does not rely on complicated integration of pre-synaptic spikes:

\begin{equation}
\label{mempot}
\tau_m\frac{dV}{dt} = -V(t) +V_{\text{rest}} + R_e CS(t) -R_i A(t) + R_p P(t).
\end{equation}

\noindent The membrane potential is controlled by three sources of current: $CS$, the 0-$1$ valued impulse train transmitted by the parallel fibers with resistance $R_e$; $A$, internally-generated inhibitory current the archive with resistance $R_e$; and, $P$, internal, excitatory current generated by the cell's intrinsic pacemaker, which we model as a Poisson impulse train with resistance $R_p$ and rate $\lambda$. When the membrane potential reaches a threshold, $V_\text{threshold}$, the potential spikes to $V_\text{spike}$ and then reset to $V_\text{hyperpolarization}$. 

\begin{table}
\begin{center}
\caption{}
  \label{cellparams}
  \begin{tabular}{ | c | c | c }
    \hline
    \emph{Parameter} & \emph{Value} \\ \hline
    $\tau_m$ & 5 ms  \\ \hline
    $V_\text{rest}$ & - 70 mV \\ \hline
    $V_\text{threshold}$ & -54 mV \\ \hline
    $V_\text{hyperpolarization}$ & -85 mV \\ \hline
    $V_\text{spike}$ & 10 mV \\ \hline
    $R_e$ & 50 $\Omega$ \\ \hline
    $R_i$ & 2.25 $\times10^6$ $\Omega$ \\ \hline
    $R_p$ & 50 $\Omega$ \\ \hline
    $\tau_\text{write}$ & 70 ms \\ \hline
    $\tau_\text{read}$ & 200 ms \\ \hline
    $AE_{\text{0},\text{write}}$ & 0 \\ \hline
    $AE_{\text{0},\text{read}}$ & 0 \\ \hline
    $AE_{\text{thresh},\text{write}}$ & 2 \\ \hline
    $AE_{\text{thresh},\text{read}}$ & 2 \\ \hline
    $R_\text{max}$ & 1 \\ \hline
    $\tau_\text{reserve}$ & 100 ms \\ \hline
    $q$ & 1.25 $\times 10^{-7}$ recorder units/ms \\ \hline
    $\sigma $ & 40 ms\\ \hline
    $\lambda$ & .3 \\ \hline
    $c$ & 3 $\times 10^{-2}$ \\ \hline
  \end{tabular}
\end{center}
\end{table}

\indent The switches on both the write and read modules are controlled by a simple exponential decay function involving activation energy :
\begin{equation}
\label{acten}
\frac{dAE_\text{module}}{dt}=\frac{1}{\tau_\text{module}}[AE_{0,\text{module}} - AE_\text{module}(t)] + CS(t).
\end{equation}
\noindent Here, $AE_\text{module}$ refers to the activation energy on the switch for either the write or read module. The time constant for each module's switch, $\tau_\text{module}$, controls the rate at which the energy passively decays to its resting value, $AE_{0,\text{module}}$, between spikes. The binary impulse train $CS(t)$ increments $AE_\text{module}$ by $1$ with each parallel fiber spike.
\newline
\indent If the pre-synaptic spikes arriving at the Purkinje cell from CS stimulation are sufficiently frequent, $AE_\text{write}$ is driven to a threshold, $AE_{\text{thresh},\text{write}}$ and the write module switches to the ON state. At this point, recorder units are released and allowed to evolve. The write module's reserve can hold a maximum of $R_\text{max} = 1$ recorder units, representing 100 $\%$ fullness. When the write module switches on, the reserve depletes exponentially with time constant $\tau_\text{reserve}$ and its refractory timer is reset so that it cannot detect the CS for a fixed interval. As soon as the reserve begins to empty, it begins to slowly repopulate at a constant rate $q$. For simplicity, we assume that the state $s$ of a released recorder unit $r$ is real-valued, so that a recorder unit can encode any time between CS onset and US onset up to some maximum granularity $\Delta t$. At each time-step, the batch of released recorder units is blurred with a Gaussian kernel of standard deviation $\sigma = 40$ ms to simulate noise.
\newline
\indent At the first US spike, the released batch of recorder units is added to a cumulative archive, unless US onset is less than 100 ms. If the first US spike occurs before this time, we assume that the batch is simply discarded since neither the Purkinje cell pause nor the overt blink can be learned for ISIs less than 100 ms. The shape of the batch, and therefore the shape of the archive, is controlled by the number of trials and the ISI (Fig. \ref{archives}). For short ISIs greater than 100 ms, the stored batch will consist mostly of recorder units approximately encoding the ISI, since noise had little time to take effect. For longer ISIs, noise overwhelms the evolution of recorder units, effectively smoothing the histogram and reducing the peak near the ISI.
\newline
\indent The read module acts in parallel to the write module. When pre-synaptic spikes are sufficiently rapid to drive $AE_\text{read}$ to a threshold, $AE_{\text{thresh},\text{read}}$, the read module switches to the ON state, and a constant fraction $c$ of recorder units are sampled from the archive. After activation, the read module's refractory timer resets so that reading is deactivated for a fixed interval. Each unit in the read batch then introduces a fixed amount of inhibitory current to the membrane at the time corresponding to its state: if read switches ON at time $t=0$ and the archive contains $n_r$ recorder units encoding state $s = t_1$, then $A(t_1) = cn_r$. Thereby, the shape of the stored archive (i.e the histogram estimate on the ISI) is directly translated into inhibitory current. The recorder units sampled from the archive during reading are permanently removed so that reading will gradually deplete the archive without compensatory addition from writing.

\section{Results}
\begin{figure}[h!]
 
\begin{subfigure}{0.5\textwidth}
\includegraphics[width=0.9\linewidth, height=5cm]{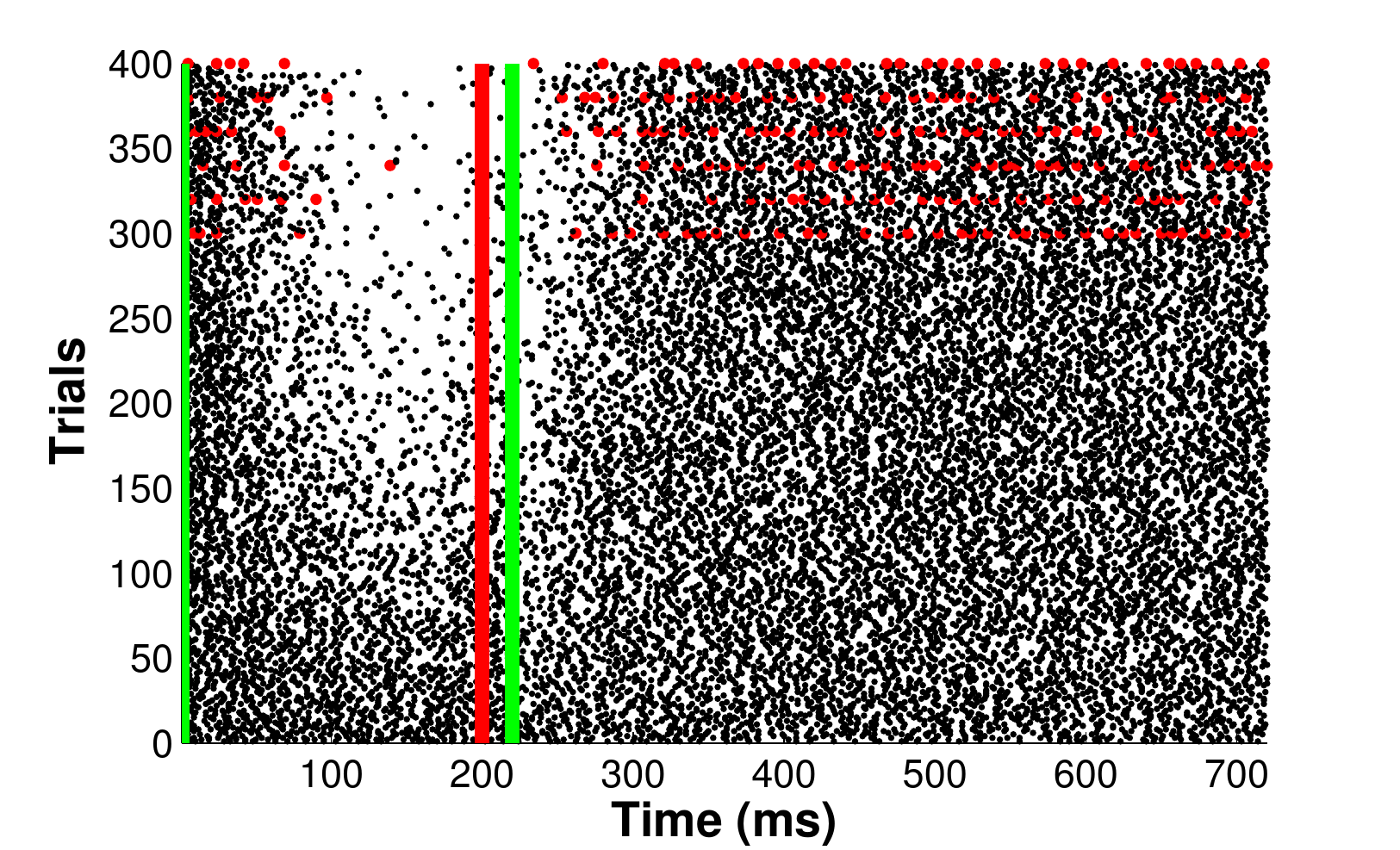} 
\caption{}
\label{sim1200msISIraster}
\end{subfigure}
\begin{subfigure}{0.5\textwidth}
\includegraphics[width=0.9\linewidth, height=5cm]{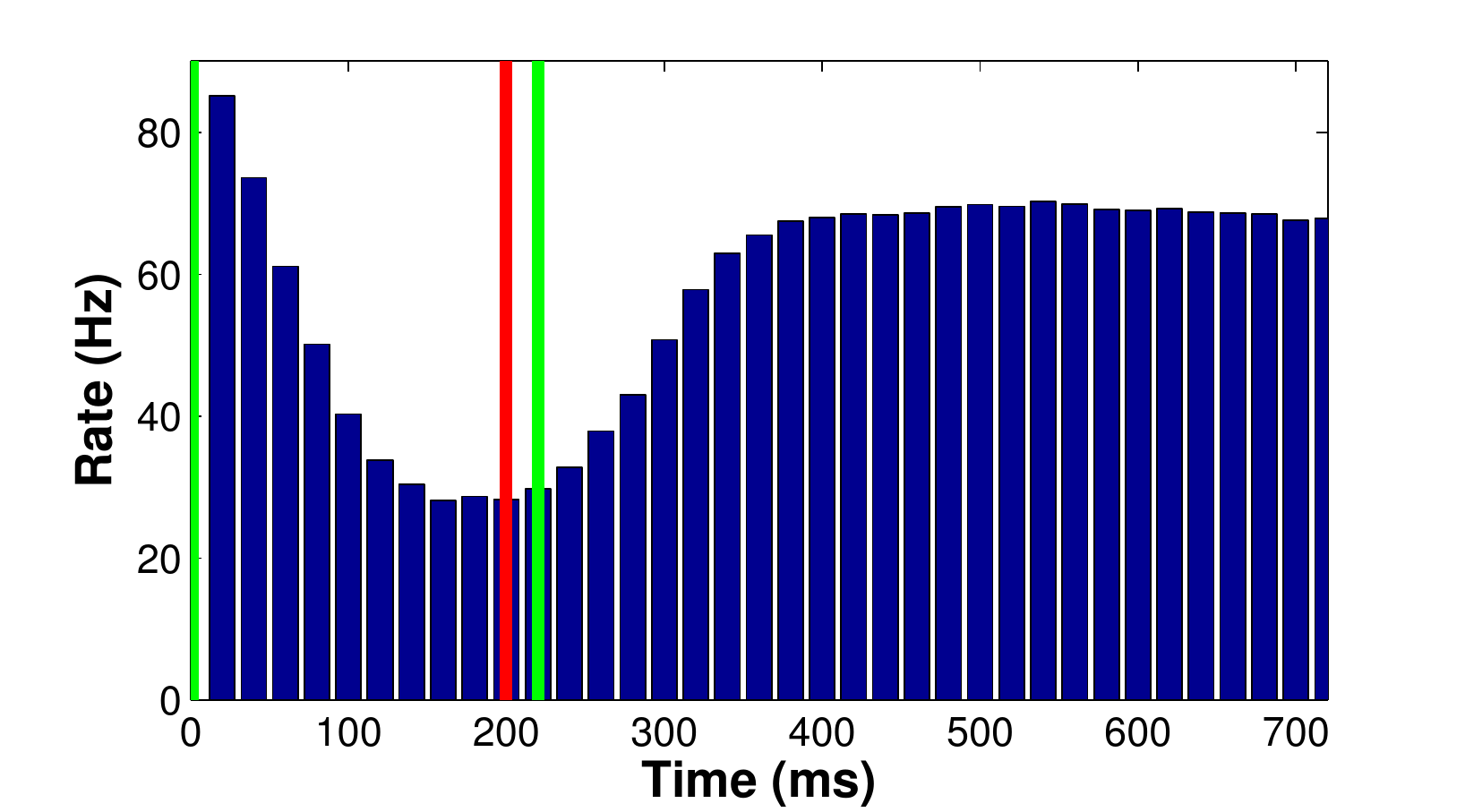}
\caption{}
\label{sim1200msISIhist}
\end{subfigure}
\begin{subfigure}{.5\textwidth}
\includegraphics[width=0.9\linewidth, height=5cm]{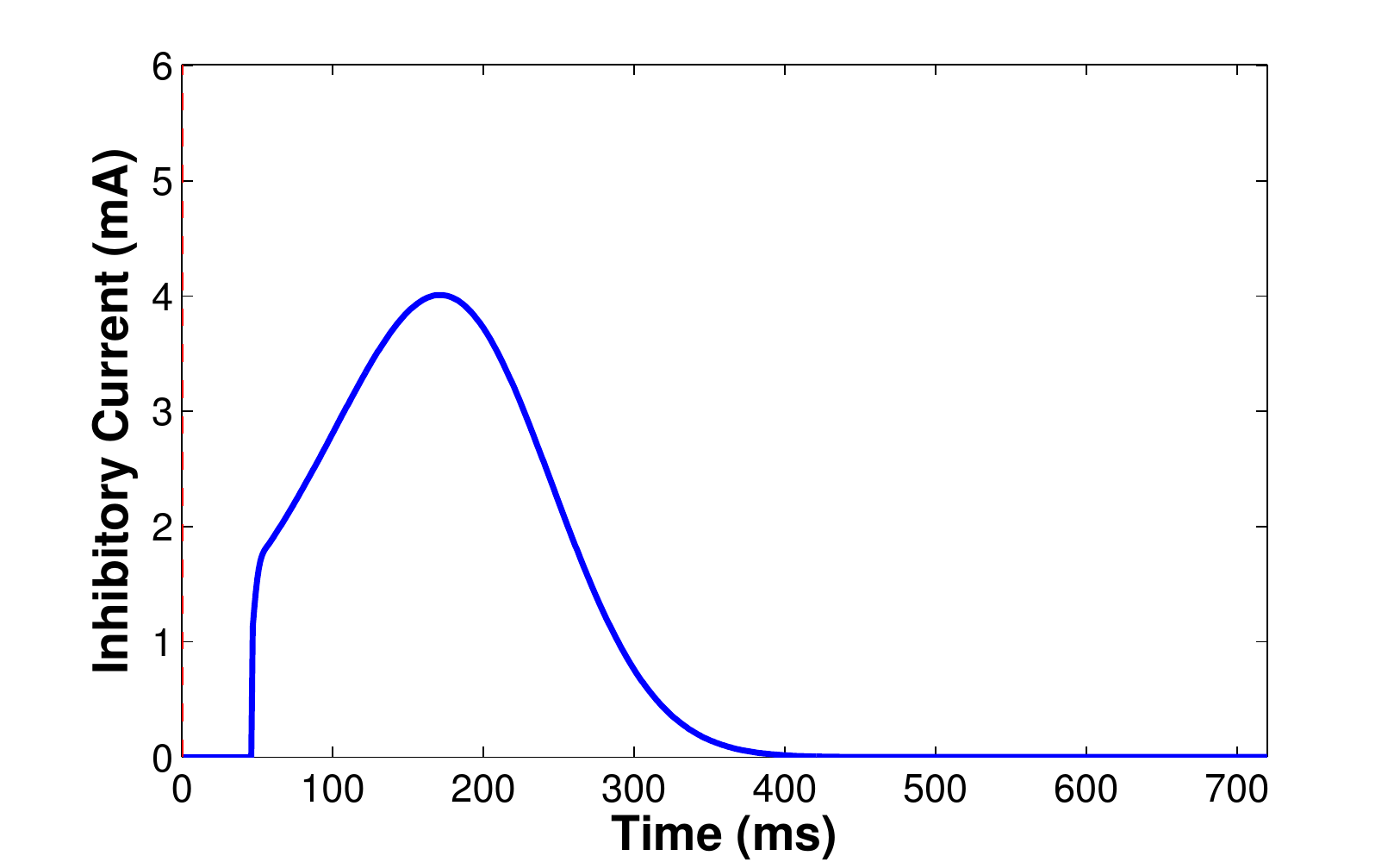}
\caption{}
\label{sim1200msISIinhibition}
\end{subfigure}
\begin{subfigure}{0.5\textwidth}
\includegraphics[width=0.9\linewidth, height=5cm]{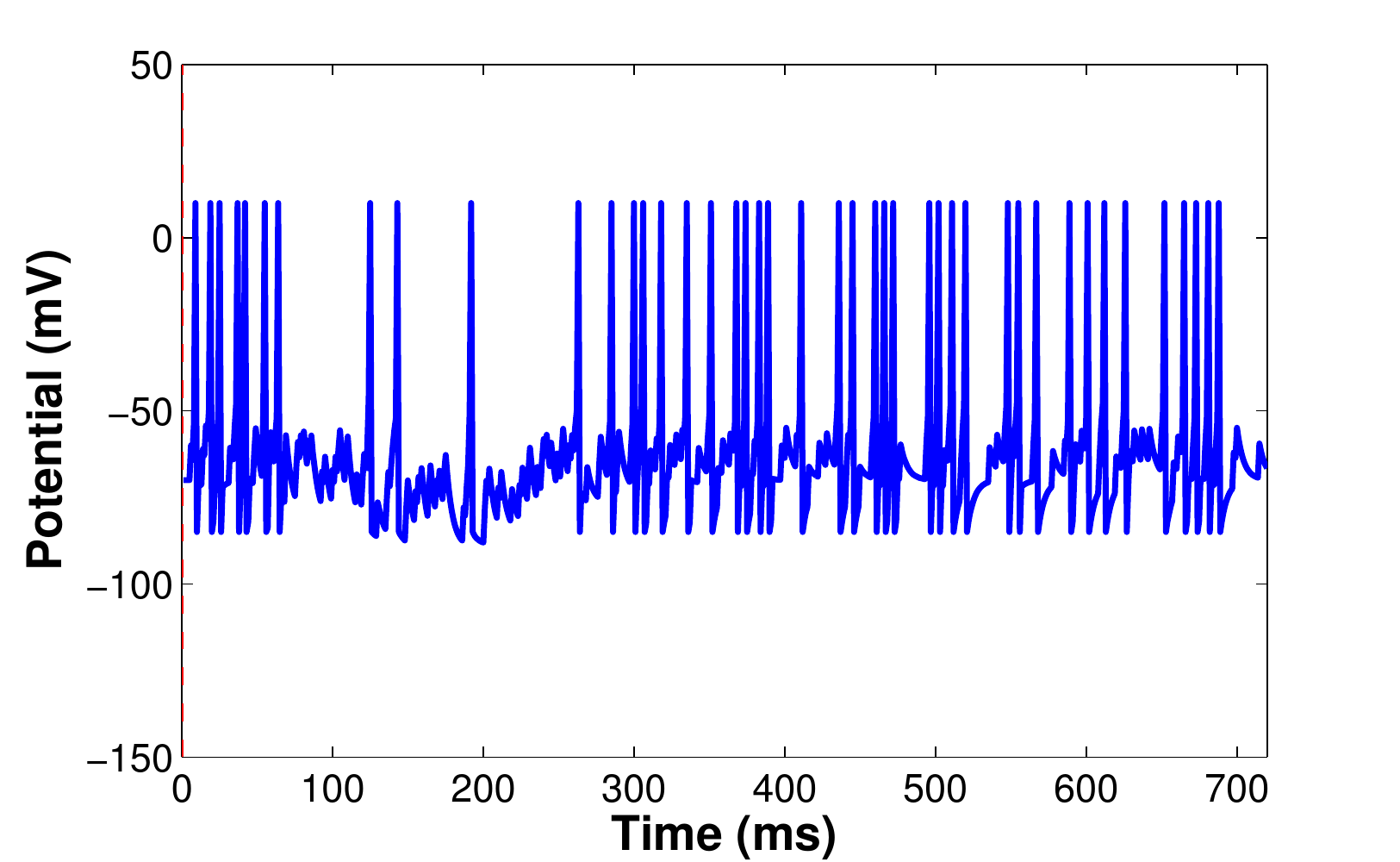}
\caption{}
\label{sim1200msISIpotential}
\end{subfigure}
 
\caption{\footnotesize \emph{Basic CR acquisition}. \emph{a)} Raster plot of spike times from Simulation 1. Green lines indicate CS onset and offset. Red line indicates US onset. Probe trials are marked by red spikes. Probe trials still produce the learned pause. \emph{b)} Peristimulus time histogram from Simulation 1. Vertical lines indicate stimulation times, as in panel \emph{a}. \emph{c)} Inhibition caused by the archive on trial 400. Note that the timecourse of inhibition mimics the shape of the archives displayed in Fig. \ref{archives1}, offset by a few tens of milliseconds. This offset is caused by the time it takes for the read module to detect CS onset and initiate hyperpolarization. \emph{d)} Actual membrane potential during interstimulus period on trial 400. Note the thinning of spikes during the ISI, indicating a well-timed CR}
\label{fig:image2}
\end{figure}

\begin{figure}[h!]

\begin{subfigure}{0.33\textwidth}
\includegraphics[width=\linewidth, height=5cm]{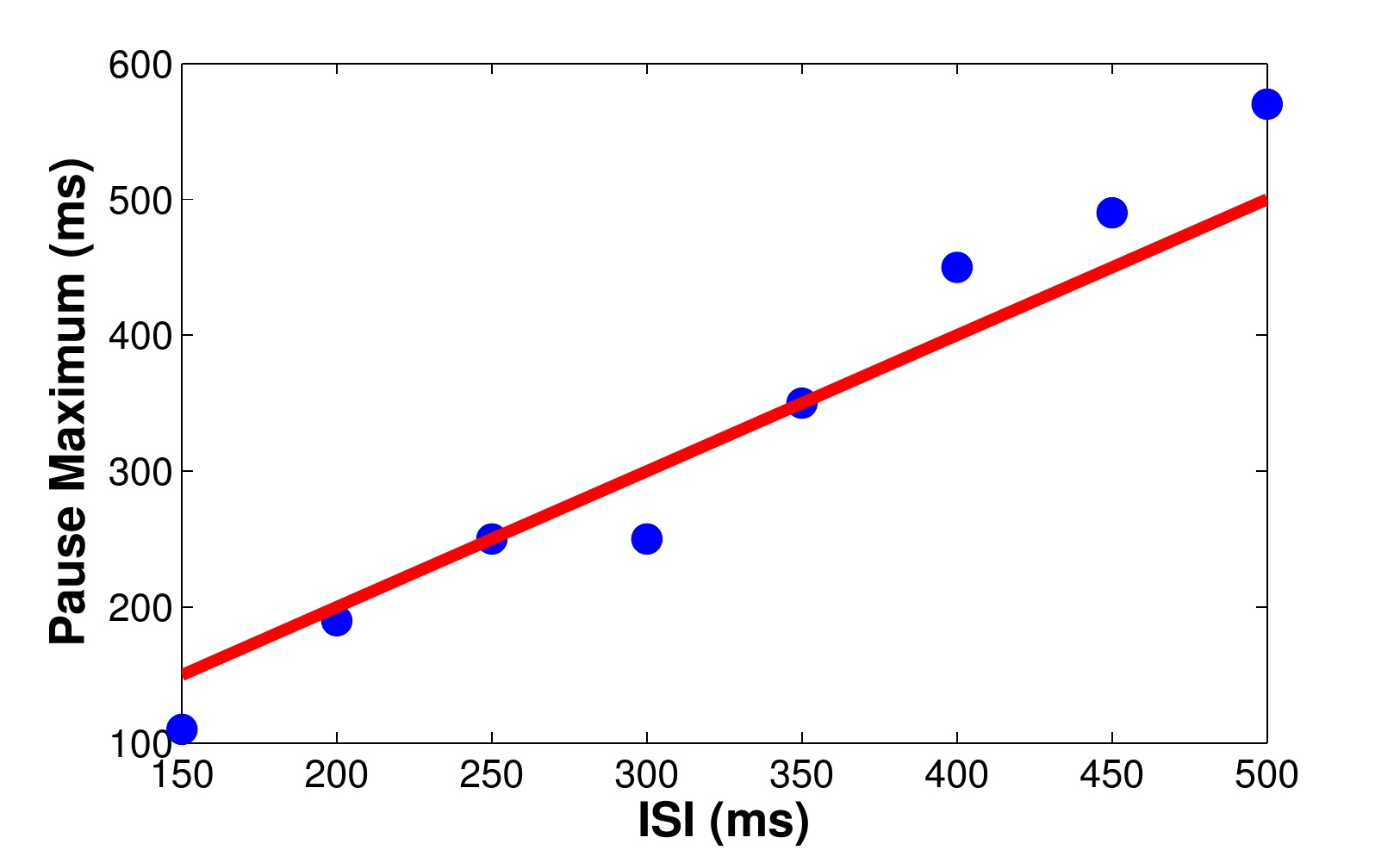} 
\caption{}
\label{sim1manyisia}
\end{subfigure}
\begin{subfigure}{0.33\textwidth}
\includegraphics[width=\linewidth, height=5cm]{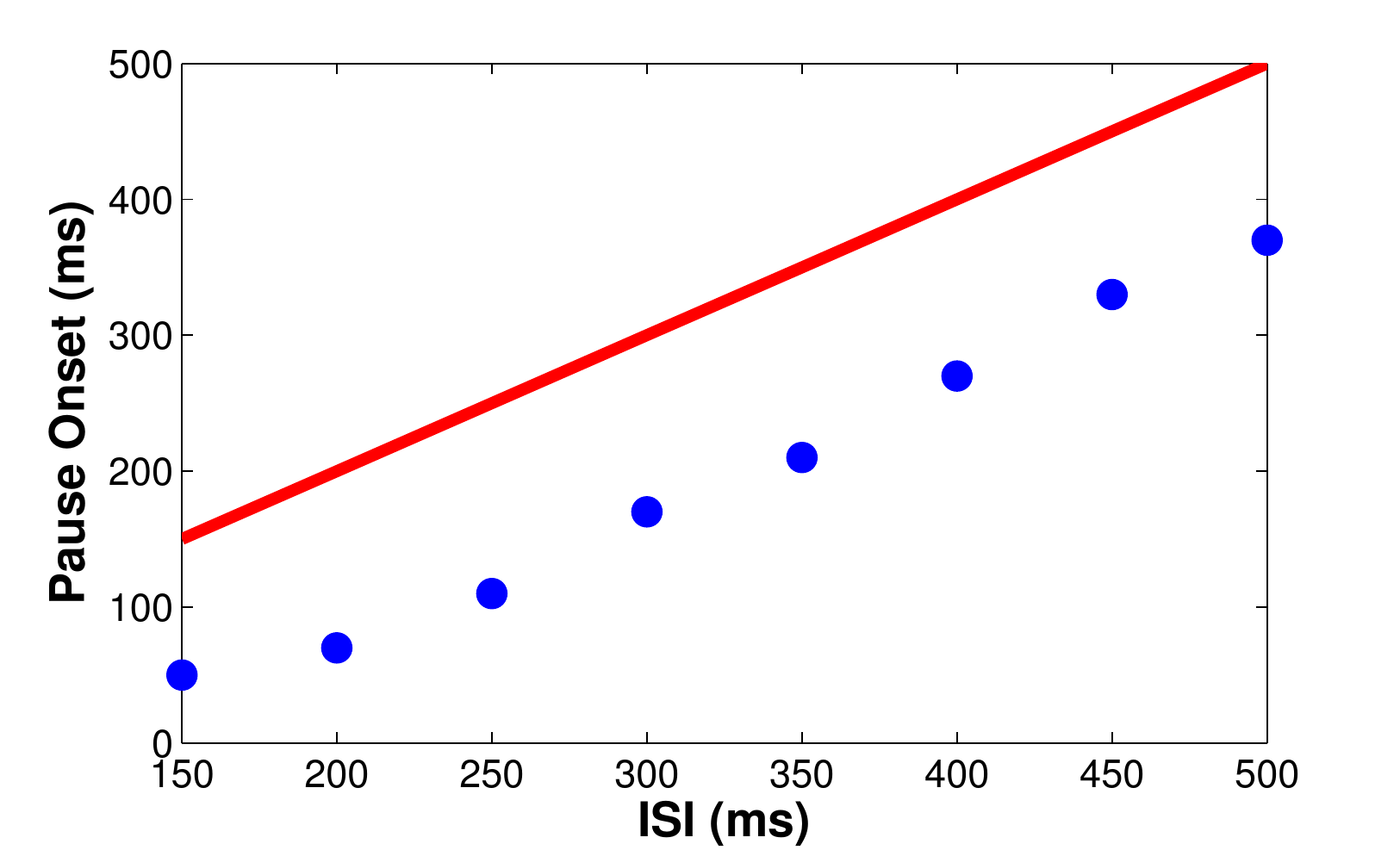}
\caption{}
\label{sim1manyisib}
\end{subfigure}
\begin{subfigure}{.33\textwidth}
\includegraphics[width=\linewidth, height=5cm]{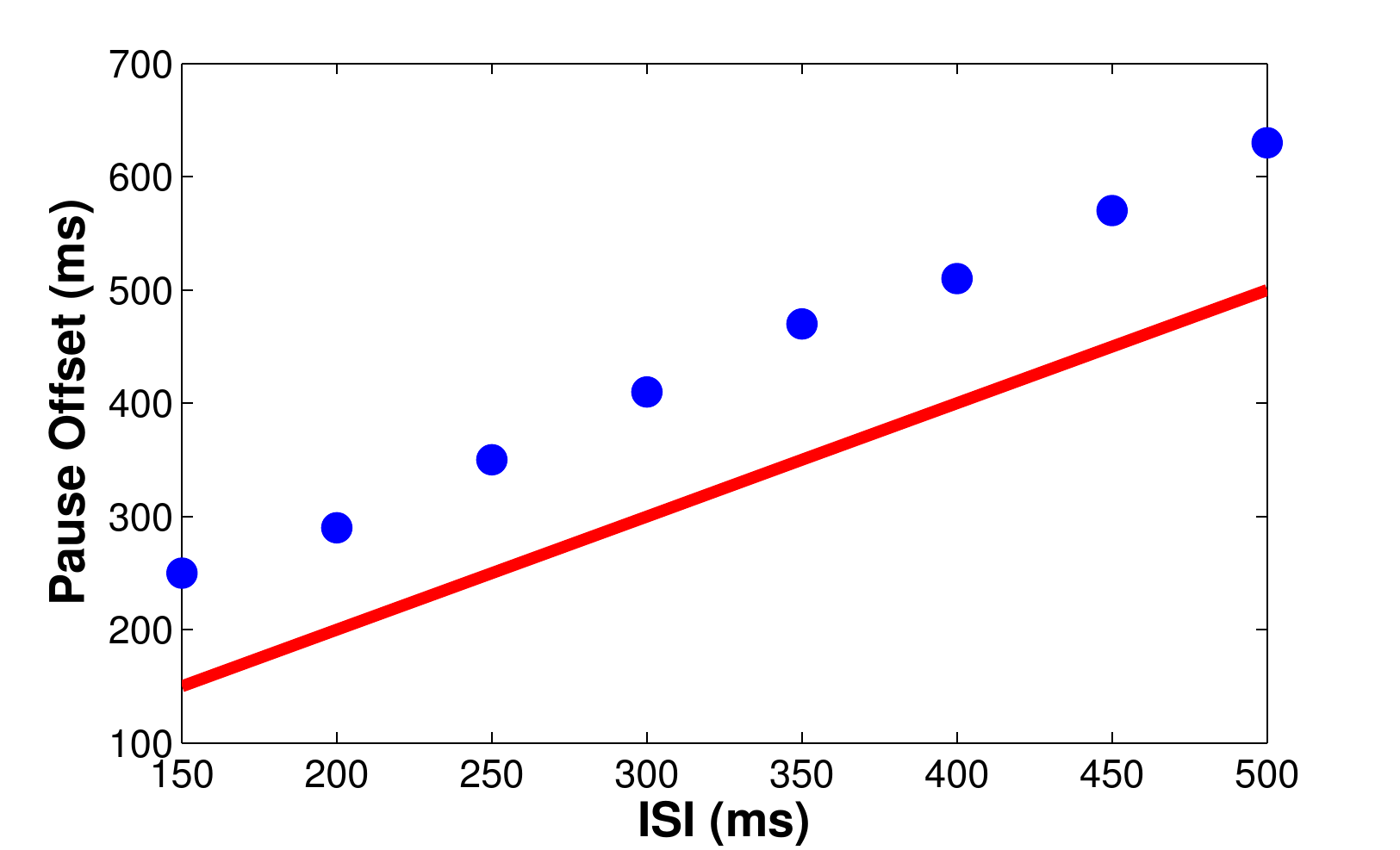}
\caption{}
\label{sim1manyisic}
\end{subfigure}
\caption{\footnotesize \emph{CR temporal features.} In these plots, blue dots indicate a temporal feature of the CR (maxima, onfsets, offsets) and the red line indicates US onset \emph{a)}. Pause maxima in Simulation 1 occurred essentially at US onset. \emph{b)} Pause onsets always precede the expected US since the archive contains some recorder units encoding low time steps. \emph{c)} Correspondingly, pause offsets always occur shortly after US onset because there are some recorder units encoding time-steps longer than the ISI.}
\label{fig:image3}
\end{figure}

\noindent \textbf{Simulation 1: } \emph{The cell learns a well-timed pause}. First, we trained the model cell on a 200 ms ISI (Fig. \ref{sim1200msISIraster}). CS stimulation consisted of a 100 Hz impulse train enduring for 220 ms and US stimulation consisted of a 500 Hz impulse train enduring for 20 ms and beginning at 200 ms. In this experiment, CS and US stimulation co-terminated. The cell was trained over the course of 400 trials with an ITI of 15 s. We declared that trial on which the average spiking rate during the ISI dropped to below 25 $\%$ of the pre-training value to be the start of the CR. In this simulation, the cell learned the pause at trial 177, which took 44.25 minutes of experimental time. Additionally, starting after 300 trials, we removed US stimulation on every 20$^\text{th}$ trial. On these probe trials (Fig. \ref{sim1200msISIraster}, red spikes), the cell still produced a pause visually indistinguishable from that of non-probe trials. A peristimulus time histogram, created by binning spikes in windows of 20 ms, averaging across trials, and smoothing with a moving average filter of span 5, is shown in Fig. \ref{sim1200msISIhist}. The inhibition created by the archive and its effect on on the cell's membrane potential are shown in Fig. \ref{sim1200msISIinhibition} and \ref{sim1200msISIpotential} respectively.
\newline
\indent Next, we trained the cell on a full battery of ISIs ranging from 150 ms to 500 ms in increments of 50 ms. All other training parameters were fixed from the first 200 ms ISI simulation. For this wide range of ISIs, the cell learned pauses with well-timed onsets (Fig. \ref{sim1manyisia}), maxima (Fig. \ref{sim1manyisib}) and offsets (Fig. \ref{sim1manyisic}). On average, pause onsets preceded US onset by 127.5 ms, pause maxima followed US onset by 7.5 ms, and pause offsets followed US onset by 110 ms. The timing of these pause features mimics those found in experimental data \citep{Johansson2014}. By design, the model cannot learn ISIs less than 100 ms. 
\newline

\begin{figure}[h!]
\begin{subfigure}{0.5\textwidth}
\includegraphics[width=0.9\linewidth, height=5cm]{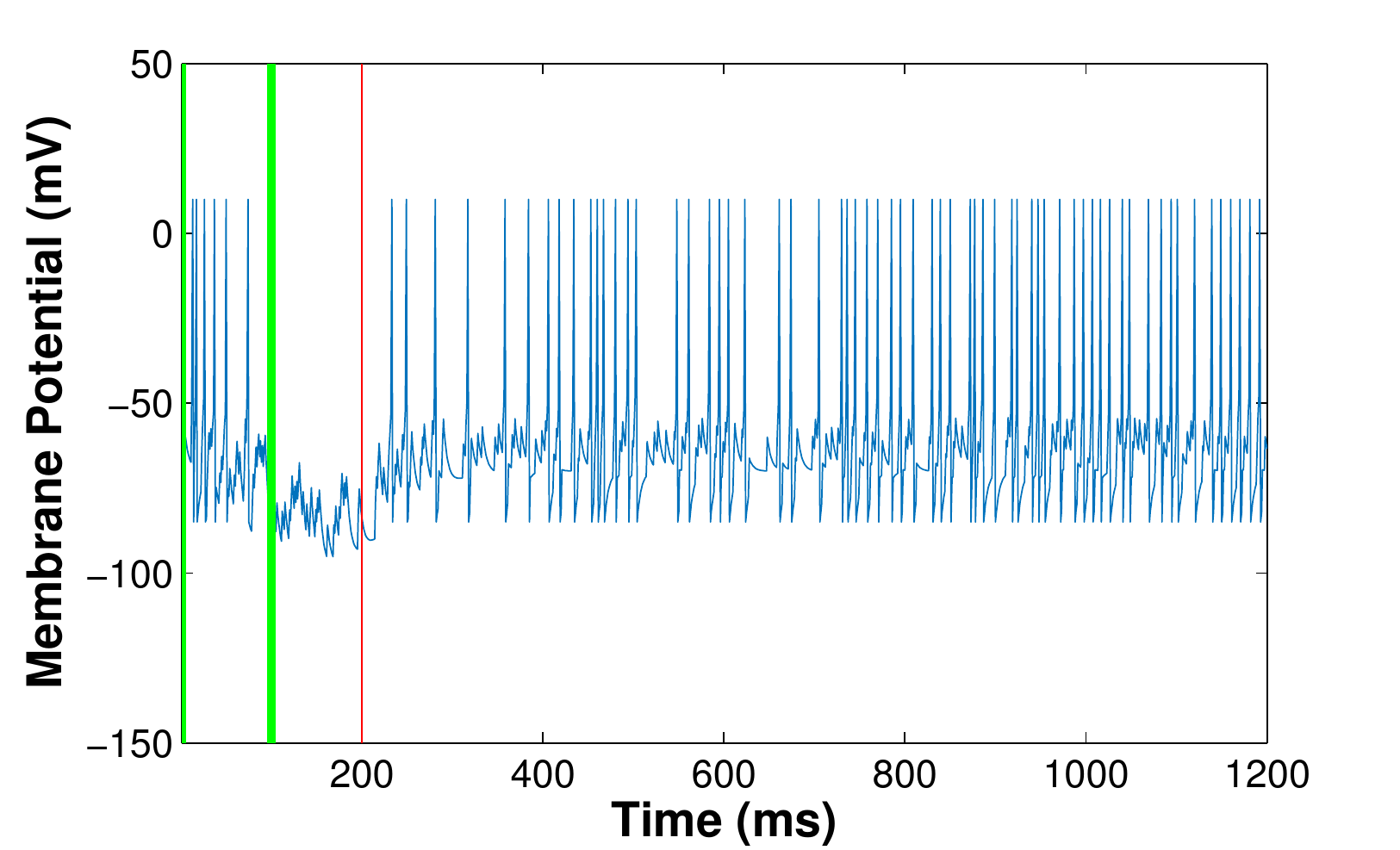} 
\caption{Probe = 50 ms, 100 Hz}
\label{sim2short}
\end{subfigure}
\begin{subfigure}{0.5\textwidth}
\includegraphics[width=0.9\linewidth, height=5cm]{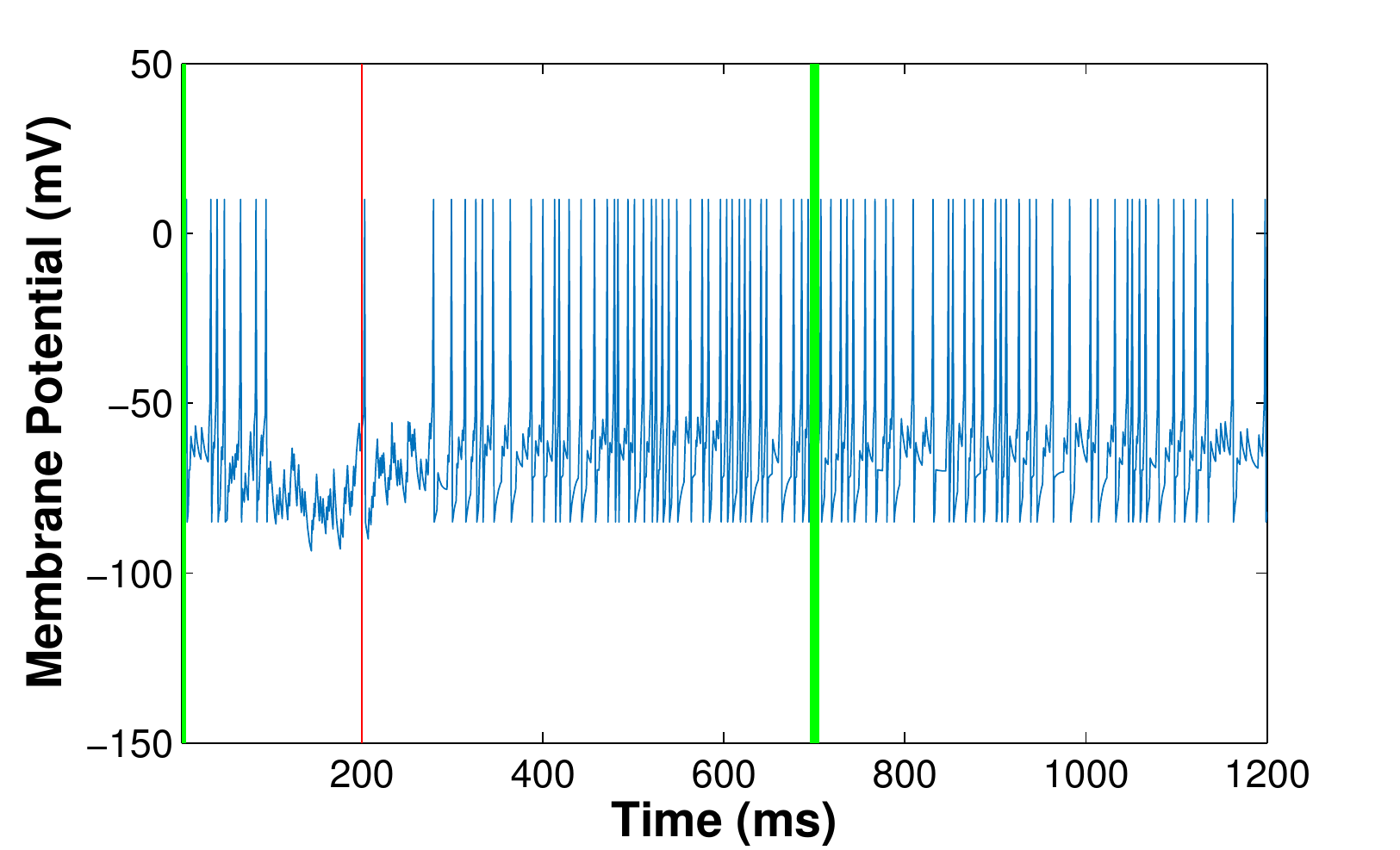}
\caption{Probe = 200 ms, 100 Hz}
\label{sim2long}
\end{subfigure}
\begin{subfigure}{.5\textwidth}
\includegraphics[width=0.9\linewidth, height=5cm]{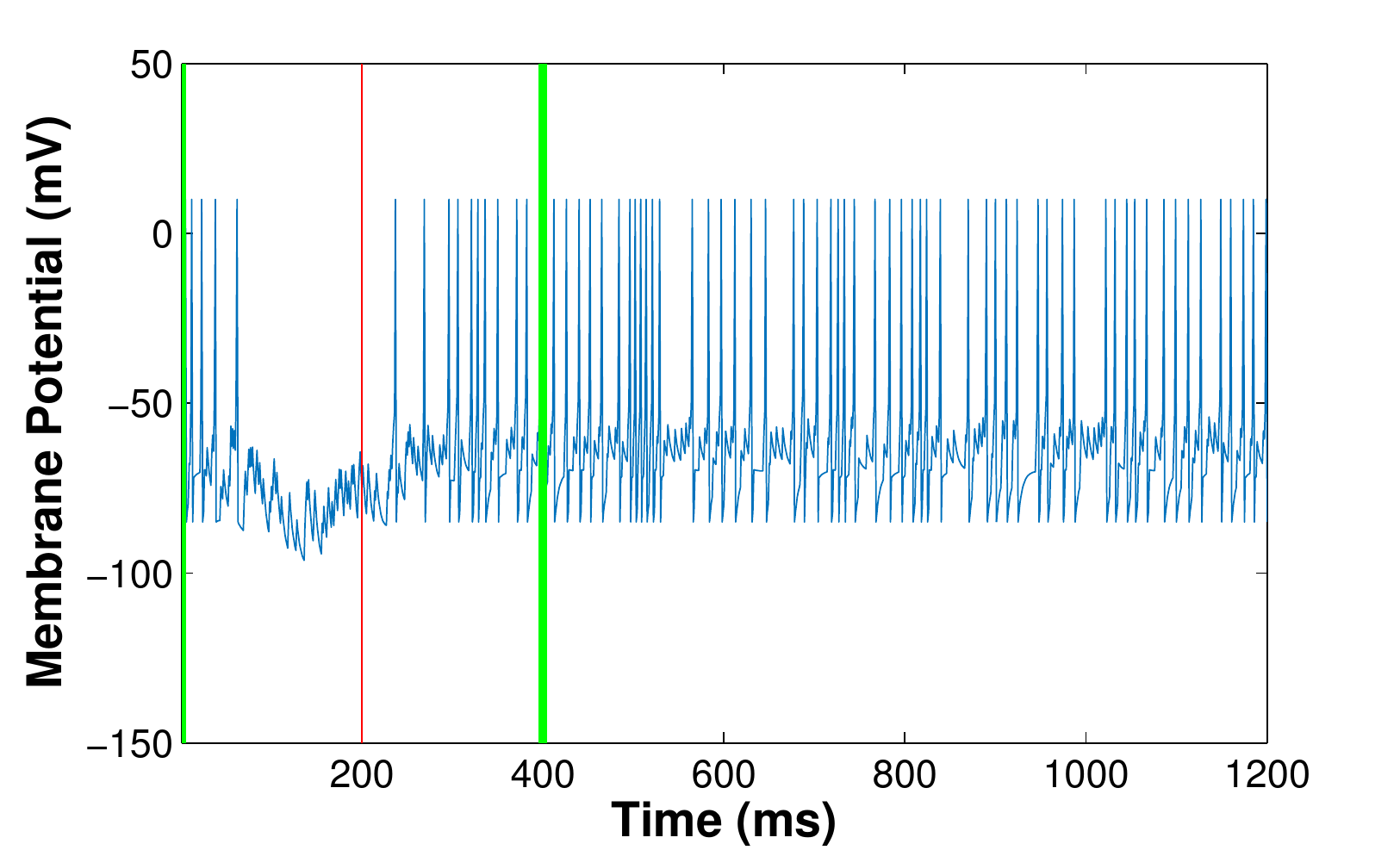}
\caption{Probe = 100 ms, 50 Hz}
\label{sim2weak}
\end{subfigure}
\begin{subfigure}{0.5\textwidth}
\includegraphics[width=0.9\linewidth, height=5cm]{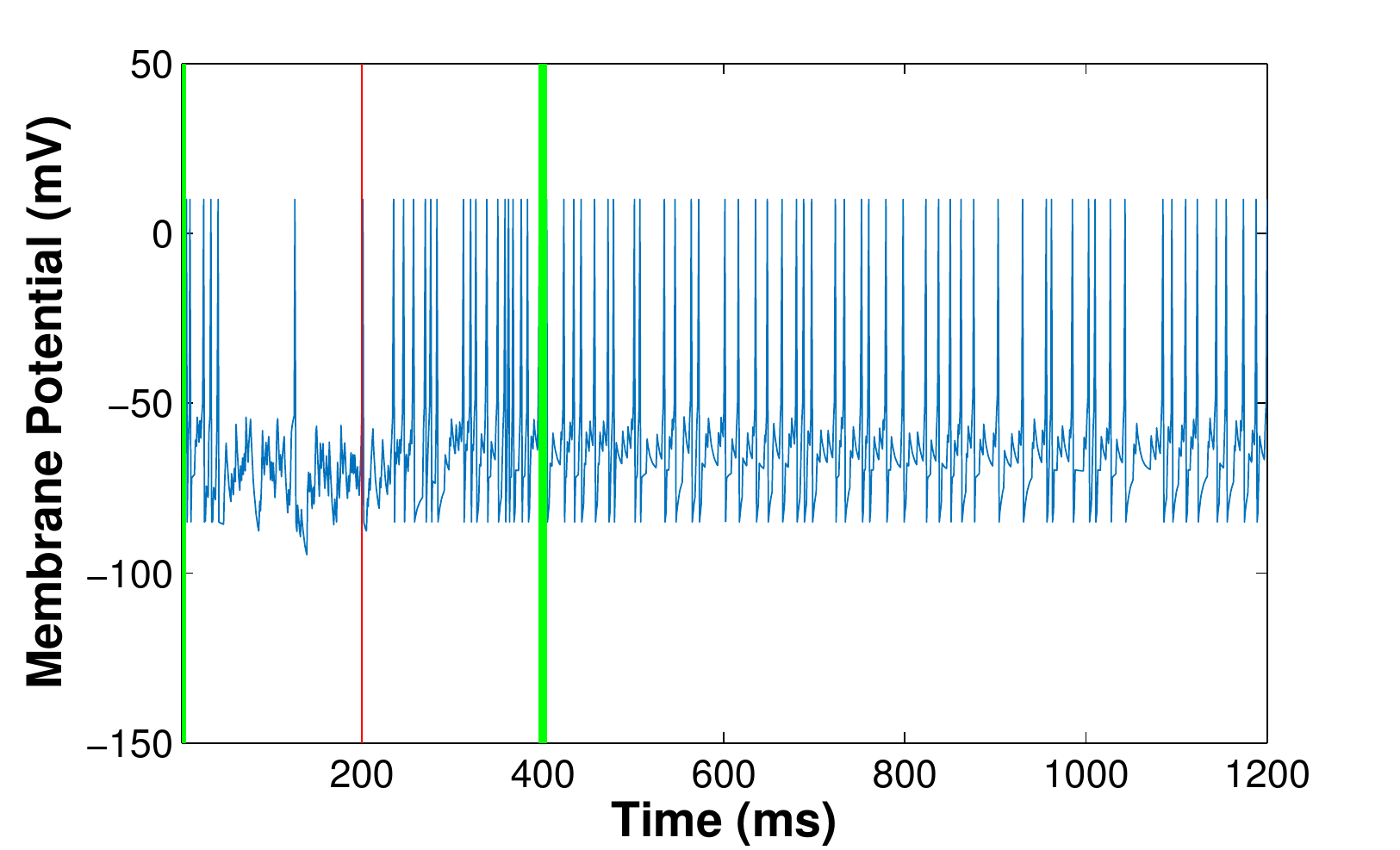}
\caption{Probe = 100 ms, 200 Hz}
\label{sim2strong}
\end{subfigure}
 
\caption{\footnotesize \emph{The CR is invariant to probe duration and frequency.} For all panels, green lines indicate CS onset and offset and red lines indicate US onset. \emph{a,b}) Changing the length of probe CS to 50 or 200 ms does not affect the temporal features of the CR. \emph{c,d)} Changing the frequency of probe CS to 50 or 200 ms does not affect the temporal features of the CR. Note that this simulation also demonstrates the CR is timed to the ISI and not the training CS's duration.}
\label{fig:image2}
\end{figure}

\noindent \textbf{Simulation 2: } \emph{The cell is invariant to probe CS}. Next, we stimulated the trained cell with probe CSs enduring for lengths different from those used during training. Like in \citet{Johansson2014}, we trained a cell with a 100 Hz 200 ms and probed it with CSs with durations 50 ms and 700 ms (Fig. \ref{sim2short}, \ref{sim2long}) and with frequencies 50 Hz and 200 Hz (Fig. \ref{sim2weak}, \ref{sim2strong}). Independent of probe CS duration and frequency, the cell still produced a well-timed CR. This invariance arises from the model's switch mechanism: a small amount of CS stimulation is sufficient to trigger the reading of the stored pause information, during which afferent parallel fiber activity has little effect. This simulation demonstrates that the model satisfies the computational constraint stipulated in the introduction: the learned pause is expressed on probe trials, independent of CS duration.
\newline

\begin{figure}[h!]
\includegraphics[width=\textwidth]{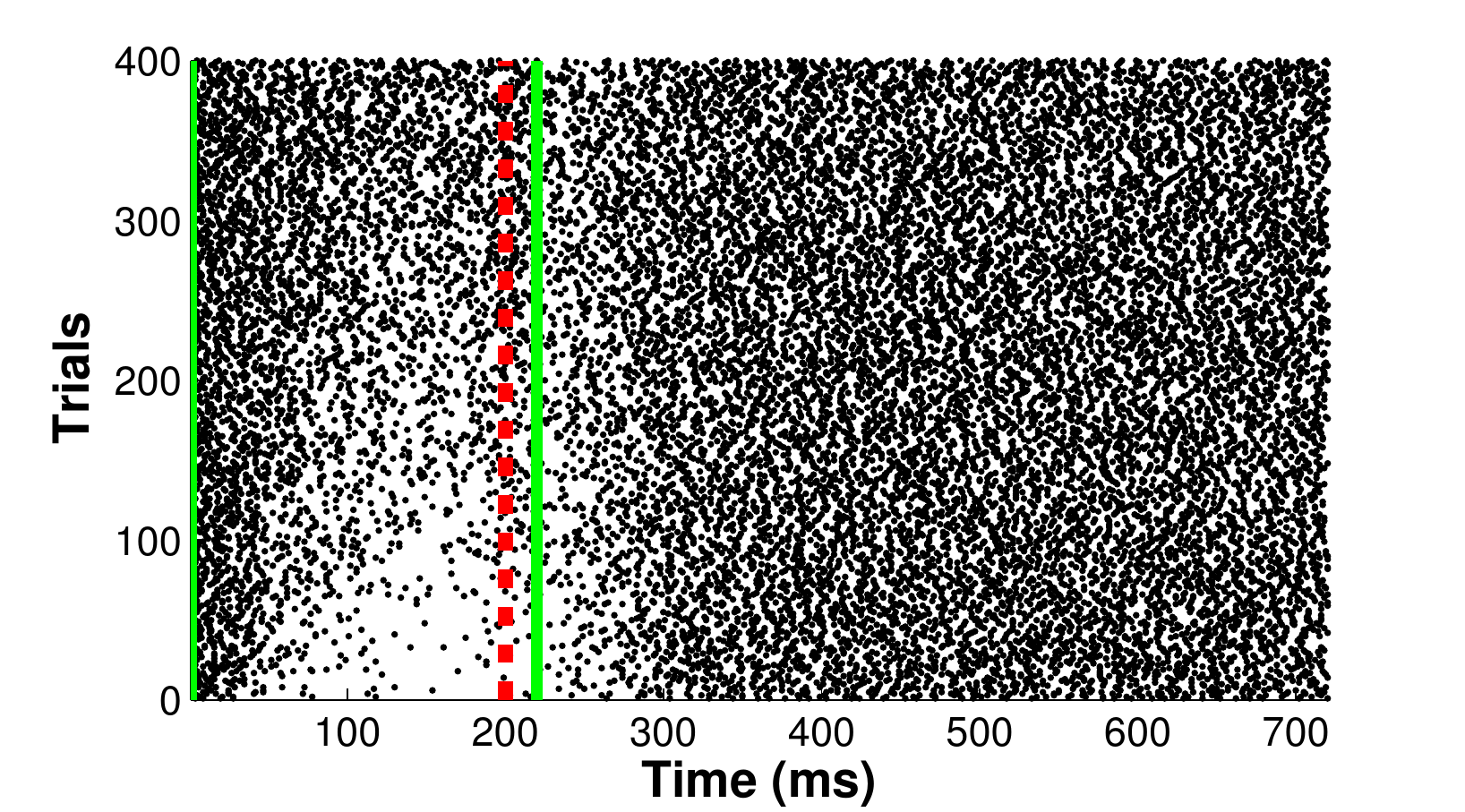}
\caption{\footnotesize \emph{CR extinction}. Green lines indicate CS onset and offset, and the red dashed line indicates expected US onset, though no US stimulation was provided in this simulation. After many CS-only trials, the CR disappears, though the archive is not empty. It has simply been reduced enough to permit tonic spiking.}
\label{ext}
\end{figure}

\noindent \textbf{Simulation 3: } \emph{The pause is extinguished over many CS-only trials}. Because the read module gradually depletes the archive, many CS-only trials will eventually extinguish the learned pause response. To demonstrate this behavior, we provided a cell previously trained on a 200 ms ISI with an additional 400 CS-only trials. The raster plot in Fig. \ref{ext} shows the gradual extinction of the pause. Within 75 trials, the cell increased to 50 $\%$ of its pre-training rate, and by trial 397, it had increased to over $90\%$. This is in line with experimental data which shows the timescale of extinction is typically less than or equal to that of acquisition.
\newline

\begin{figure}[h!]

\begin{subfigure}{0.5\textwidth}
\includegraphics[width=\linewidth, height=5cm]{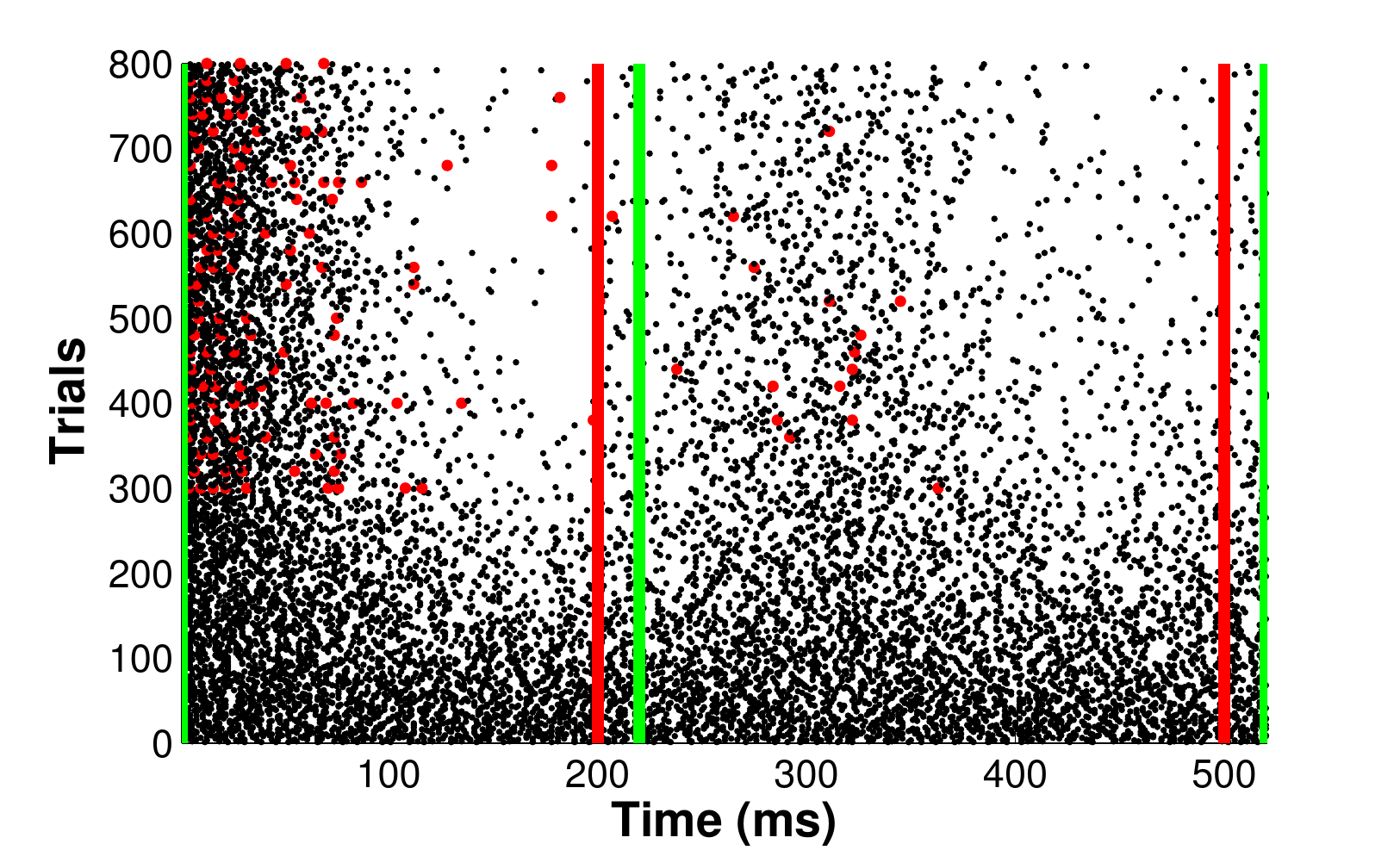}
\caption{}
\label{}
\end{subfigure}
\begin{subfigure}{0.5\textwidth}
\includegraphics[width=\linewidth, height=5cm]{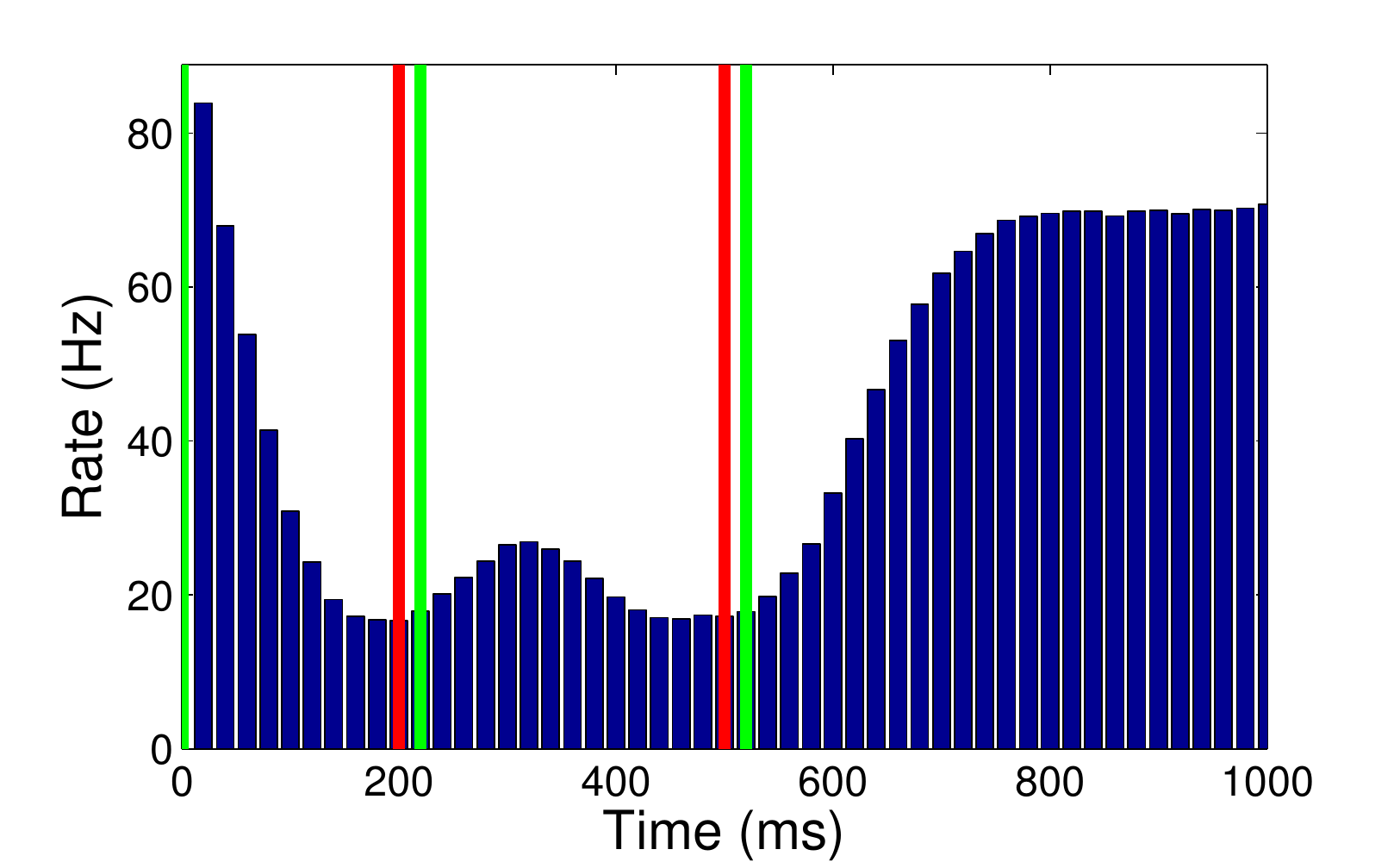}
\caption{}
\label{sim4hist}
\end{subfigure}
\caption{\footnotesize \emph{A bimodal pause}. \emph{a)} Green lines indicate CS onset and offset, the red line indicates US onset, and red spikes indicate probe trails. In this simulation, the cell learned two ISIs, one timed to 200 ms and another timed to 500 ms. The CR consists of two pauses with an intervening resumption of spiking between 200 and 400 ms. \emph{b)} The bimodal CR is evident from a raster plot created in the same manner as described in Simulation 1. Presumably, this paradigm could be generalized to more than 2 pauses, as long as the modes of the archive are well-separated enough to allow for spiking resumption between the constituent pauses.}
\label{bimodal}
\end{figure}

\noindent \textbf{Simulation 4: } \emph{The cell can learn multiple pauses}. \citet{Jirenhed2007} and \citet{Johansson2014} found that interleaving trials with different ISIs teaches the Purkinje cell two CRs so that, given one probe CS, the cell will pause twice. Our model Purkinje cell reproduces this behavior (Fig. \ref{bimodal}) in the case of interleaved 200 ms and 500 ms ISI trials with a 30 s ITI. However, in this simulation, both pauses were weaker and took more time to express (197 trials vs. Simulation 1's 177 trials) using the same detection criterion. This is because only half of the trials contribute to the learning of each ISI, and those trials which do not contribute to one ISI still consume the archive by engaging the read module. As a result, experimental time to acquisition is more than twice that of the original case (98.5 minutes vs. 44.25 minutes), adjusted for ITI. 
\newline

\noindent Simulations 1-4 have been performed in the real Purkinje cell. For simulations 5 and 6, we ran simulations which, to our knowledge, have not been performed experimentally. 
\newline 

\begin{figure}[h!]
\begin{subfigure}{0.5\textwidth}
\includegraphics[width=\linewidth, height=5cm]{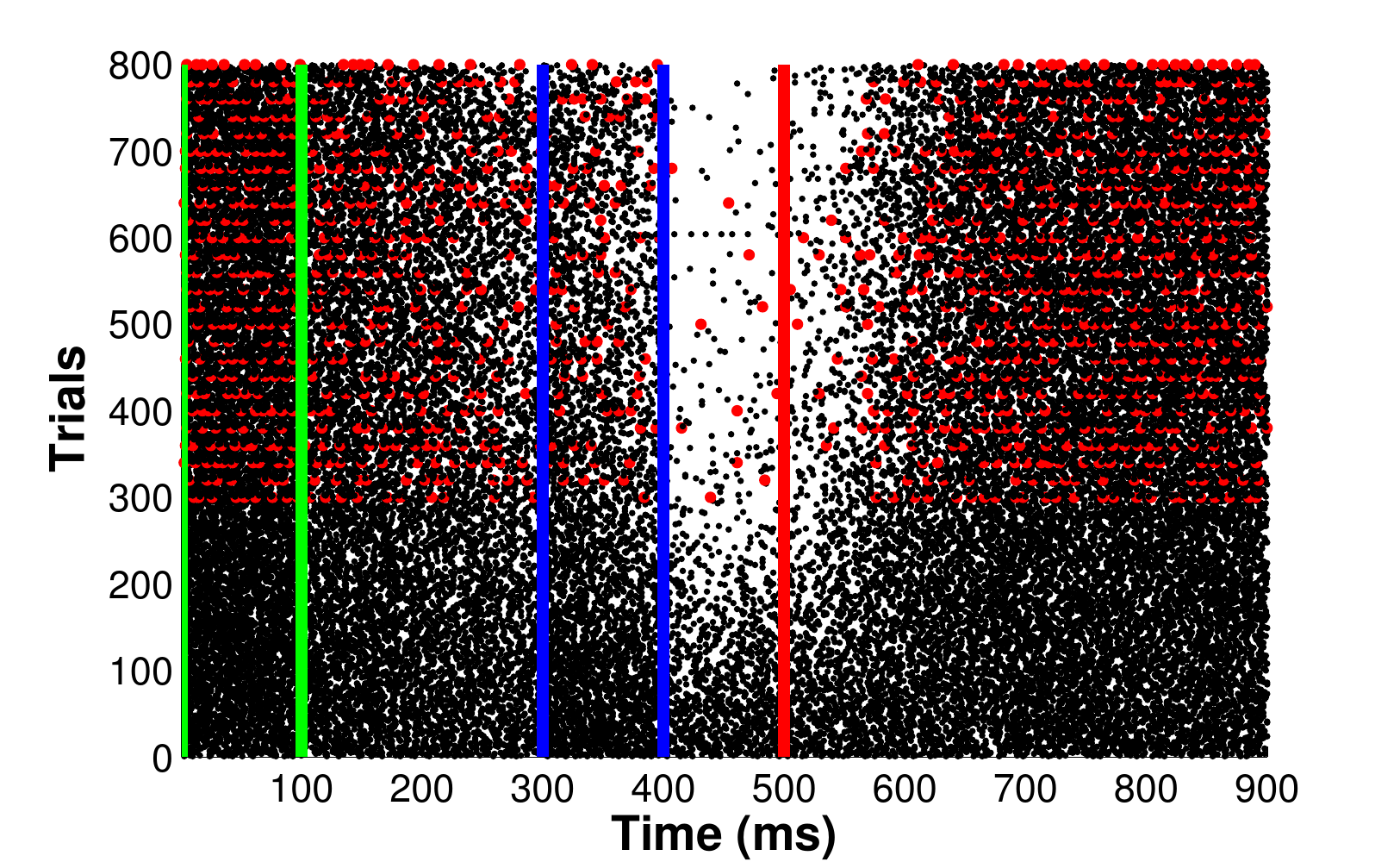}
\caption{}
\label{sim52cs}
\end{subfigure}
\begin{subfigure}{0.5\textwidth}
\includegraphics[width=\linewidth, height=5cm]{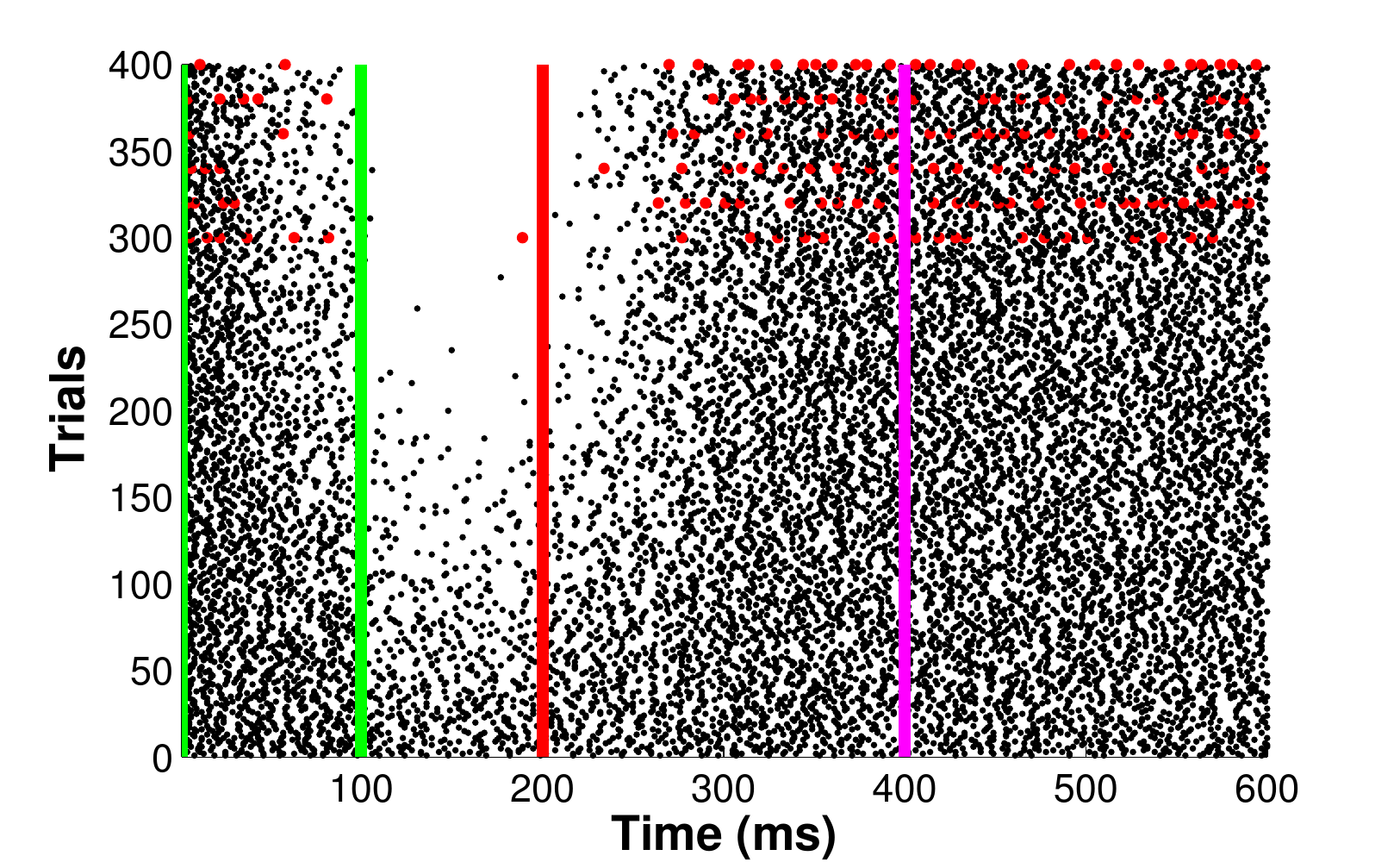}
\caption{}
\label{sim52us}
\end{subfigure}
\caption{\footnotesize \emph{With 2 CSs or USs per trial, only one CR develops.} \emph{a)} Green lines indicate the onset and offset of the first CS; blue lines indicate the onset and offset of the second CS; red line indicates US onset. A CR appears beginning at around 400 ms and terminates shortly after the US. If the cell had learned the second ISI, the pause would appear and disappear much earlier in the trial, since the second ISI is only 200 ms. \emph{b)} Red line indicates the first US onset; magenta line indicates second US offset. Spiking resumes around 300 ms since only the shorter ISI is learned.}
\label{2cs2us}
\end{figure}

\noindent \textbf{Simulation 5: } \emph{2 CSs or 2 USs}. First, we trained the model cell with two CSs on each trial. One CS began at $t=0$ and ended at $t=100$ ms; another began at $t = 300$ ms and ended at $t=400$ ms. US onset was at 500 ms and lasted for 20 ms. 
Hence, the cell was effectively exposed to two ISIs on each trial, 500 ms and 200 ms. Like previous simulations, both CSs were 100 Hz and the US was 500 Hz. The ITI was 15 s. Whereas interleaving two ISIs across trials was found to produce two pauses in Simulation 4, providing the cell with two CSs on each trial produced only one in an experiment lasting 800 trials. (Fig. \ref{sim52cs}). The cell only learns the longer ISI since the refractory period on the write switch prevents it from detecting the second CS. 
\newline
\indent Next, we stimulated the cell with two USs on each trial, one at 200 ms and another at 400 ms, both lasting for 20 ms at 500 Hz. The simulation lasted for 400 trials with an ITI of 15 s. The cell learned one pause, timed to the shorter ISI (Fig. \ref{sim52us}). As in the double CS experiment, the cell is becomes insensitive to further US stimulation after the first US onset. 
\newline

\begin{figure}[h!]
\includegraphics[width=\textwidth]{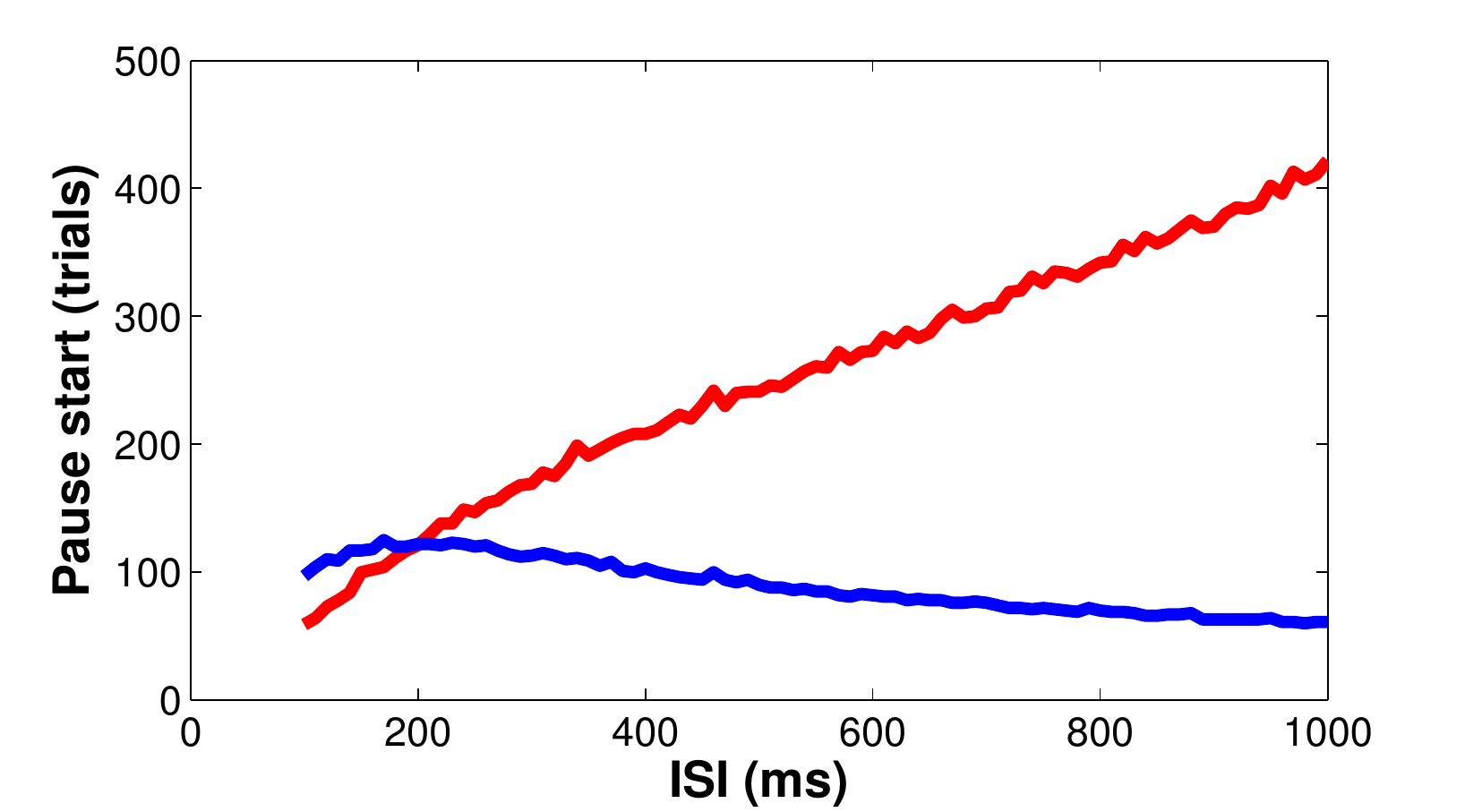}
\caption{\footnotesize \emph{ITI affects trials to acquisition.} Red line indicates trials to acquisition in experiment in which ISI varied between 100 and 1000 ms and ITI was fixed at 15 s; blue line indicates trials to acquisition in experiment in which the ISI varied over the same range, but ITI was fixed at 80 times the ISI. The lines intersect when the ITI is 15 s. Otherwise, the red line increases linearly due to the effect of noise on long ISIs and the inability of the reserve to completely replenish in 15 s. The blue line, on the other hand, is constant, since the increased noise of long ISIs is compensated for an increased ITI.}
\label{ratio}
\end{figure}

\noindent \textbf{Simulation 6: } \emph{Trials to acquisition depends on ISI and ITI}. The number of trials until pause acquisition is affected by several factors. First, longer ISIs are harder to learn since noise during writing tends to flatten the histograms added to the archive, which consequently takes more trials to build up enough inhibition to cancel spiking. Short ISIs yield highly peaked histograms and therefore speed up learning. Further, long ITIs tend to speed up learning since they allow the reserve to replenish fully and release a maximum number of recorder units during subsequent reading. Short ITIs, on the other hand, do not allow for full replenishment, so that fewer recorder units are released and the archive grows more slowly.
\newline
\indent To investigate how these factors precisely affect our model cell, we ran two parallel experiments. In one experiment, we fixed the ITI at 15 s and trained the cell on ISIs ranging from 100 ms to 1000 ms in steps of 10 ms. For the other experiment, we used the same range of ISIs, but also varied the ITI so that ITI/ISI $= 80$ for each ISI. In the first experiment, trials to acquisition increased monotonically and approximately linearly with ISI (Fig. \ref{ratio}, red line). For short ISIs noise in recorder unit evolution smooths the archive and delays learning. Further, the relatively short ITI prevents the write module's reserve from replenishing fully between trials. However, when the ITI/ISI ratio was held constant, the hindering effect of noise due to increasing ISI was canceled out by the accelerating effect of the increased ITI, resulting in a flat trials-to-acquisition curve (Fig. \ref{ratio}, blue line). This effect has not yet been tested in the Purkinje cell, though it has been observed in other behavioral experiments \citep{Gibbon1977}. The finding illustrated in Fig. \ref{ratio} suggests that trials to acquisition $T$ is given roughly by $T = 8000 ($ISI/ITI$)$. 

\section{Discussion}
This purely formal model recapitulates the behavior of the Purkinje cell during eyeblink conditioning. Most importantly, it learns a pause response with critically-timed onset, maximum and offset. Further, the model obeys the two desiderata stipulated in the introduction: the POT representation makes no use of granule cells and is internal to the POT; and, pause learning and expression machinery is gated by a switch. Obeying these criteria saves this model from some of the crucial limitations afflicting previous POT models of the Purkinje cell. Additionally, the model makes several predictions:
\newline
\begin{itemize}
\item If the Purkinje cell is stimulated with multiple CSs on each trial, it will only learn the longest ISI. 
\item If the Purkinje cell is stimulated with multiple USs on each trial, it will only learn the shortest ISI. 
\item The number of trials until pause acquisition is a constant times the ISI/ITI ratio
\end{itemize}

\indent \citet{Johansson2014}, \citet{Chen2014} and \citet{Ryan2015} may call for a fundamental re-imagining of the neural basis of learning and memory. The model outlined above is a computational proof-of-principle for a learning mechanism inspired by this work, in the case of eye-blink conditioning. Formally, the model in many ways resembles earlier population models of the Purknje cell learning mechanism, though here a population of neurons has been replaced with a population of abstract ``recorder units''. The model makes several straightforward predictions, which can be tested electrophysiologically. Most importantly, a complete understanding of the Purkinje cell will require insight from molecular biology, in order to illuminate the structure of the learning mechanism this paper attempts to formally describe. 
\bibliographystyle{apalike}

\end{document}